\documentclass[a4paper,fleqn,usenatbib]{mnras}

\usepackage{newtxtext,newtxmath}

\usepackage[T1]{fontenc}
\usepackage{ae,aecompl}

\usepackage{graphicx}	
\usepackage{amsmath}	

\title[Best Distances to 402 Galactic Novae]{Comprehensive Catalogue of the Overall Best Distances  and Properties of 402 Galactic Novae}

\author[B. E. Schaefer]{
Bradley E. Schaefer$^{1}$\thanks{E-mail: schaefer@lsu.edu},
\\
$^{1}$Department of Physics and Astronomy, Louisiana State University, Baton Rouge, Louisiana, 70820, USA\\
}

\pubyear{2021}

\begin{document}
\label{firstpage}
\pagerange{\pageref{firstpage}--\pageref{lastpage}}
\maketitle

\begin{abstract}

I derive the overall best distances for all 402 known galactic novae, and I collect their many properties.  The centrepiece is the 74 novae with accurate parallaxes from the new {\it Gaia} data release.  For the needed priors, I have collected 171 distances based on old methods (including expansion parallaxes and extinction distances).  Further, I have collected the V-magnitudes at peak and the extinction measures, so as to produce absolute magnitudes at peak and then derive a crude distance as a prior.  Further, I have recognized that 41 per cent of the known novae are concentrated in the bulge, with 68 per cent of these $<$5.4$\degr$ from the galactic centre, so the 165 bulge novae must have distances of 8000$\pm$750 parsecs.  Putting this all together, I have derived distances to all 402 novae, of which 220 have distances to an accuracy of better than 30 per cent.  I find that the disc novae have an exponential scale height of 140$\pm$10 pc.  The average peak absolute V-magnitude is -7.45, with an RMS scatter of 1.33 mag.  These peak luminosities are significantly correlated with the decline rate ($t_3$ in days) as $M_{V,peak} = -7.6 + 1.5 \log(t_3 / 30)$.  The huge scatter about this relation masks the correlation in many smaller datasets, and makes this relation useless for physical models.  The bulge novae are indistinguishable from the disc novae in all properties, except that the  novae with red giant companion stars have a strong preference for residing in the bulge population.
 
\end{abstract}




\section{INTRODUCTION}

Since the earliest days, astrophysics has had the difficult primary task of measuring the distances to objects of all classes.  For the classical novae{\footnote{The classical novae (CNe) are cataclysmic variables (CVs) in binaries with a relatively ordinary companion star spilling mass into an accretion disc around a white dwarf, which display runaway thermonuclear explosions on the surface of the white dwarf (Warner 2008).  These nova eruptions have outburst amplitudes $\gtrsim$8 mags, rise times of days to weeks, and durations of weeks to many months (Payne-Gaposchkin 1964).  An important class of CNe is the recurrent novae (RNe), which are ordinary novae that happen to have recurrence time-scales faster than a century  (Schaefer 2010).}, measuring accurate distances has also long been an enterprise of the highest importance.  Starting around the 1920s, novae were being distinguished as a separate class, distinct from other transients, with the nova distances bound up in the debates over the separate existence of galaxies as well as the separate existence of supernovae.  Starting soon after the 1901 eruption of GK Per, a small number of novae had measured distances from the expansion parallax method, wherein the expanding nova shell had some vague edge matched to some poorly-defined radial velocity from an eruption spectrum.  For many decades, the distances to individual novae were few and unreliable, with the best being the relative distances for the novae seen in globular clusters, the Magellanic Clouds, and the Andromeda Galaxy.  Even by the time of Payne-Gaposchkin's great treatise in 1964, little else was known.

Starting around the 1980s, a wide variety of distances for individual novae were put forth, all with poor reliability.  A common on-going analysis of new novae would use some measure of the interstellar extinction, somehow calibrated, to derive distances, but such measures were uniformly poor and the calibrations had huge uncertainties.  And measures of the observed peak magnitudes (and related quantities) were used with a wide variety of poorly calibrated correlations to get distances, but even the calibration novae displayed huge scatter, so any pretense for accuracy was not justified (Schaefer 2018).  The best of the measures were from expansion parallaxes, while an impressive paper by Downes \& Duerbeck (2000) used the {\it Hubble Space Telescope} ({\it HST}) to image 30 novae, collect all prior shell size data, and came up with 29 expansion parallax distances.  Around this time, attempts were made to use novae to independently derive the distances to nearby galaxies and the Hubble Constant (Jacoby et al. 1992).

Starting around a decade ago, newly measured nova distances were being published with substantially better reliability.  For the extinction distances, \"{O}zd\"{o}nmez et al. (2018) systematically collected nova extinction measures and made good averages, and then used infrared sky surveys showing the tip of the red giant branch stars to provide a good calibration of the extinction as a function of distance along sightlines close to each nova.  Parallaxes have always been the best method to get reliable distances to stars, and nova parallaxes started to be produced, first for 4 novae with the {\it HST} by Harrison et al. (2013), then 3 novae with the first data release of {\it Gaia} (Ramsay et al. 2017), then I used the {\it Gaia} Data Release number 2 (DR2) parallaxes to report 41 accurate nova distances (Schaefer 2018).

The natural follow-up to the DR2 list of nova distances is to use the {\it Gaia} third data release (DR3) to construct a set of improved nova distance for many more novae.  This is a primary goal of this paper.  The new {\it Gaia} data provides good-accuracy distances to 74 novae.  In addition, DR3 provides low-accuracy distances (often really just lower limits) for 121 novae.  To realize these distances, the correct Bayesian calculations require the use of prior information, with such never having been assembled previously for any set of novae.  This prior information needs the assembly of all prior published distances from all the many old methods found throughout the literature.  Further, the prior information needs the magnitudes and extinctions for the many {\it Gaia} novae.  So the correct use of the good and the poor {\it Gaia} parallaxes requires a large scale data mining program.

Once we have the idea of making a large catalog of the best distances for the {\it Gaia} novae, a natural extension is to get the best distances for $\it all$ known galactic novae.  For the novae not in {\it Gaia} DR3, all we have is a collection of non-parallax data.  For roughly a third of the novae, the only useful information is the galactic position.  For the bulge novae, this turns out to give accurate and reliable distances.  For the disc novae, the position alone can provide distance constraints with the uncertainty of order 3$\times$ from the central estimate, and that is adequate for many purposes.  So my program is extended to provide the best possible distances for all galactic novae.

The first step of the program is to construct a complete census of known galactic novae for peaks with all times before middle 2022.  This list was constructed from partial lists in Duerbeck (1987), the Catalog and Atlas of Cataclysmic Variables (CVCat, Downes et al. 2001, updated to 2005), the International Variable Star Index{\footnote{https://www.aavso.org/vsx/} (VSX) of the American Association of Variable Star Observers, and the up-to-date list{\footnote{https://asd.gsfc.nasa.gov/Koji.Mukai/novae/novae.html} of K. Mukai (NASA HEASARC).  In most cases, the status as a nova is clear from the light curve and spectra, however, roughly a dozen systems are poorly observed so that the nova status is questionable, and for this I have followed the always excellent judgement of H. Duerbeck in his very useful catalog (Duerbeck 1987).  For inclusion in my list, I required good evidence that the transient was indeed a classical nova.  My list does not include the red novae (like V838 Mon and V1309 Sco), the symbiotic novae (like PU Vul and V1016 Cyg), the X-ray novae (like V404 Cyg and V616 Mon), nor systems that show only dwarf nova events (like WX Cet and WZ Sge), because these eruptions are all morphologically greatly different, the systems have different evolution, and are dominated by different mechanisms.  In the end, I have 402 systems that are confidently galactic novae.

\section{NOVA DISTANCES FROM ALL THE OLD METHODS}

The olden standard for nova distances has been the expansion parallax.  Schaefer (2018) used {\it Gaia} DR2 parallaxes and found that the real accuracy has a one-sigma error bar of 0.95 mag in the distance modulus.  Here, I have collected 38 expansion parallax distances as previously collected in Schaefer (2018) and \"{O}zd\"{o}nmez et al. (2018).

A traditional astronomy method for getting nova distances is to somehow measure the extinction from the intervening interstellar medium (ISM), then to calibrate the extinction as a function of distance along the line of sight to the nova.  The olden extinction distances were always poor.  A useful new tool has been the reddening maps showing $E(B-V)$ for the entire line of sight through our galaxy, all with high accuracy for lines-of-sight within 5 arc-minutes of any target, as based on the far-infrared dust emission (Schlafly \& Finkbeiner 2011).  Recently, a good advance has been made by \"{O}zd\"{o}nmez et al. (2018), where the extinction as a function of distance is calibrated from the brightness and color of red clump stars observed by several deep infrared sky surveys.  \"{O}zd\"{o}nmez et al. reports on 81 reddening distances, including 29 lower limits on distances.  Schaefer (2018) found that these reddening-distances have a one-sigma uncertainty of 1.01 mag in the distance moduli.

Here, I introduce another method based on extinction.  The idea is that the total extinction throughout the Milky Way along every line of sight is known from infrared mapping (Schlafly \& Finkbeiner 2011), so for systems with substantially less than the maximum, we can quantitatively estimate the nova distance.  For example, V4742 Sgr has the maximum possible extinction of 5.65 mag, while its measured $E(B-V)$ is 1.5$\pm$0.2.  With the majority of the line-of-sight extinction occurring within the region around the centre of our galaxy, V4742 Sgr must be substantially closer than 8000 pc.  To be quantitative, I have modeled the Milky Way dust as the usual axisymmetric exponential disc, with a scale height of 100 pc out of the galactic plane, and with a scale length of 3200 pc in its galactocentric radius distribution (Li et a. 2018).  From an integral within this model, I calculate the dust column density as a function of distance from Earth, and scale this by the maximum $E(B-V)$ for the entire column through our Milky Way.  The observed extinction can then be matched to distance, and this can be converted to a distance modulus.  For V4742 Sgr, the observed extinction corresponds to a distance of 3120$\pm$380 pc, or a distance modulus of $\mu$=12.47$\pm$0.26.  The formal error bars are likely much too small, partly because the real dust in our galaxy is much more clumpy than in the standard model.  I will adopt the same one-sigma uncertainty of 1.01 mag, as in the previous paragraph.  The results from this method are superseded by the better calibration of \"{O}zd\"{o}nmez et al. (2018).  The only novae for which this method provides useful constraints are KY Sgr, V4742 Sgr, V4744 Sgr, V5586 Sgr, and V1662 Sco.

A physics method for nova distances is to get a blackbody distance to the companion stars.  I have collected 7 blackbody distances from Schaefer (2010), Salazar et al. (2017), and Shara et al. (2017a).

Another physics method is to model the eruption light curve of an individual nova, scale the time to a universal law, so as to derive the peak absolute magnitude, which then gives the distance modulus.  I. Hachisu (University of Tokyo) and M. Kato (Keio University) are the only modelers with this method, from which I have collected 68 nova distances as reported in Hachisu \& Kato (2021) and references therein.  Schaefer (2018) found that these measures have a one-sigma uncertainty of 0.76 mag in the distance moduli.

The {\it HST} has been able to measure good parallaxes for four of the nearest and brightest novae; V603 Aql, DQ Her, GK Per, and RR Pic (Harrison et al. 2013).  Schaefer (2018) found these four measures to be substantially poorer than the quoted error bars, with the average one-sigma uncertainty being 0.37 mag in the distance moduli.

A unique method for measuring a nova distance has been presented by Sokolowski et al. (2013) for the RN T Pyx, wherein the light echo as reflected by the previously ejected nova shells is used similarly to the expansion parallax.  

Two novae are known to appear inside galactic globular clusters, with the nova positions being sufficiently close to the cores so that the membership in the cluster is not in any doubt.  The old nova T Sco is in the globular M80, while Nova 1938 Oph is in M14.  The distances to the host globular clusters are taken from the {\it Gaia} EDR3 parallaxes for over 600 member stars (Vasiliev \& Baumgardt 2021).

These distances are collected into Table 1.  The first column gives the nova designation, with the novae listed in the standard order as in the General Catalog of Variable Stars (GCVS).  The second column gives the reported old distances, in parsecs.  The third column gives the associated distance modulus and its assigned one-sigma uncertainty.  The last column gives the method for the old distance, with 38 expansion parallaxes, 87 distances and limits based on extinction, 7 blackbody distances from the companions, 32 distances based on the physics models of Hachisu \& Kato, 4 parallax distances from $HST$, 1 light echo distance for T Pyx, and 2 distances for the host globular clusters.  For all of the old methods, I have a total of 171 distances, of which 29 are lower limits on the distance.

\begin{table}
	\centering
	\caption{Compilation of Nova Distances from Old Methods (full table with 171 novae is in on-line supplementary material)}
	\begin{tabular}{lrrrrr}
		\hline
		Nova & D (pc) & $\mu_{old}$ (mag)  & Method \\
		\hline
OS And	&		7300	&		14.32	$\pm$	0.76	&	Model	\\
CI Aql	&		3900	&		12.96	$\pm$	1.01	&	Extinction	\\
V500 Aql	&		5900	&		13.85	$\pm$	0.95	&	Expansion	\\
V603 Aql	&		249	&		6.98	$\pm$	0.37	&	$HST$ $\varpi$	\\
V1229 Aql	&		2400	&		11.90	$\pm$	0.95	&	Expansion	\\
...   &		&		&		\\
PW Vul	&		1800	&		11.28	$\pm$	0.95	&	Expansion	\\
QU Vul	&		3140	&		12.48	$\pm$	0.95	&	Expansion	\\
QV Vul	&		2700	&		12.16	$\pm$	0.95	&	Expansion	\\
V458 Vul	&	$>$	6000	&	$>$	13.89	$\pm$	1.01	&	Extinction	\\
V459 Vul	&		3800	&		12.90	$\pm$	0.76	&	Model	\\
		\hline
	\end{tabular}
	
\end{table}

\section{DISTANCES BASED ON THE PEAK MAGNITUDES}

Despite the historical confusions, there is good information on nova distances from the peak magnitudes, and this method can be applied to most old novae.  Indeed, for many novae, the peak magnitude provides the only specific information on $D$.  Even for those novae with {\it Gaia} parallaxes, the distances from the peak magnitudes will form the primary basis for the priors as used in the Bayesian calculation of the parallax-distances. 

For my plan of getting the best distance information for all 402 galactic novae, I need to collect the peak magnitudes and extinctions for all the novae that were observed at maximum.  More specifically, I need the observed peak $V$-band magnitude, $V_{peak}$, as well as the color excess, $E(B-V)$.  These will be used to calculate the distance modulus, $\mu_{peak}$.  

In bureaucratic mode, we could just look up the peak magnitudes in various catalogs, but such would be poor for the majority of the novae.  A primary problem is that the usual tabulations often are listing only the maximum observed brightness, which might or might not be at the real peak, or worse will list some speculative peak magnitude based on extrapolation.  Another big problem is that the peak magnitudes are often in a wide variety of band passes other than $V$-band, with the applicable band being often unstated, whereas some correction is needed to get to $V_{peak}$.  These problems can be confidently solved for over a hundred of the brightest novae, with these being well observed and reported in the literature.  The best source is Strope, Schaefer, \& Henden (2010) for the 93 all-time (before 2010) best observed novae, because this work went back and pulled out all the light curves, from all the published literature, from the entire AAVSO database, plus extensive measures of my own from the Harvard plates.  Another reliable source is Duerbeck's catalog (Duerbeck 1987) for all novae up until 1987, as he also went back to the entire literature plus going back to the original photographic plates of old, and his evaluations are excellent.  Another reliable source is C. Payne-Gaposchkin's great treatise (Payne-Gaposchkin 1964), also with references to the full literature and a deep knowledge of the Harvard plates.  Even with these excellent compilations, the majority of the novae still have either unlisted or questionable peak magnitudes.  At this time, the only way to produce reliable measures of $V_{peak}$ or its limit is to go through the entire literature and AAVSO database exhaustively for roughly 300 novae.

To get a useable distance modulus, we also need a reliable measure of extinction.  For this, the literature contains useful measures of $E(B-V)$, or its equivalent, for only the brighter novae, and these are often contradictory and always scattered by much more than the quoted error bars.  \"{O}zd\"{o}nmez et al. (2018) compiled an exhaustive and excellent summary of the scattered literature to make judicious evaluations for the $E(B-V)$ of 178 galactic novae.  This provides the foundation for my tabulation of $E(B-V)$.  But we still need some sort of an estimate for the remaining 224 galactic novae.  For this, a wonderful tool is the all-sky catalog of $E(B-V)$ through the entire line of sight through our Milky Way galaxy (Schlafly \& Finkbeiner 2011).  This serves to put an approximate-but-strict upper limit on $E(B-V)$.  For some novae, this upper limit serves as an adequate constraint, while many other of the faintest novae away from the galactic plane must have their extinction close to the limit (as the nova is certainly far past all galactic dust).  For the roughly 200 novae without useful limits from \"{O}zd\"{o}nmez or Schlafly \& Finkbeiner, the only solution is to go through the literature and data for effective evidence on the extinction.  I have also pulled out good measures of extinction by extracting the $V_{peak}$ and $B_{peak}$ from the magnitudes scattered through the literature (see below).

While doing these exhausting and exhaustive searches of the literature and databases, I have also tabulated the best information on a variety of other nova properties.  From my new light curve compilations, I have classified 265 light curves into the classes S, P, O, C, D, J, and F (Strope et al. 2010).  These light curve classes are strongly correlated with a variety of nova properties.  From the literature (primarily the Strope catalog and the Duerbeck catalog) and from my own compilations of the light curves, I have measured both $t_2$ and $t_3$ (the number of days for the nova light curve to decline from peak magnitude by 2 and by 3 magnitudes).  From all this, I have only been able to extract 262 and 252 reliable measures of $t_2$ and $t_3$ respectively.  I have also compiled lists of the spectral classifications for 259 nova, with the classes being Fe {\rm II}, He/N, and Hybrid, as well as whether it is a Neon Novae.  I have also compiled a list of the shell expansion velocity for 216 galactic novae, as measured by the FWHM of the H$\alpha$ line.  This measure changes substantially with the development over time, and the literature often only reports the FWHM of other hydrogen lines or the FWZI of the Balmer lines, so this compilation has inhomogeneities at perhaps the 30 per cent level.  My list of spectral classes and FWHM values is just a large extension of the list presented in Pagnotta \& Schaefer (2014).  Schaefer (2022) has already compiled a list of the orbital periods for 156 galactic novae, with this list including 49 newly discovered orbital periods and the list is 2$\times$ longer than previous lists.  Finally, I have also compiled the galactic coordinates for all 402 novae, mainly from the VSX database.


Importantly, I have been careful for distinguishing whether the true peak was observed.  For inclusion in my list of measured $V_{peak}$, I require some suitable evidence that the brightest observed magnitude is close in time to the real peak.  This evidence is usually an observed rise to maximum, or a useful closely-spaced pre-eruption limit, or a spectroscopic `age'.   When multiple measures on the night of peak are available, I use the nightly average.  When I determine that the nova light curve does not reliably represent the true maximum, I list the $V_{peak}$ as a limit.

The light curves and the tabulated peak magnitudes are usually in the $V$-band, however, 102 nova from before 1973 have only $B$-band light curves, almost always based on the archival Harvard plates.  The spectral sensitivity of these plates is essentially identical to that of the Johnson $B$-band, so there is no color-term for the reported magnitudes.  The peak magnitudes for novae are always in the bright regime, where the comparison stars are accurately on the modern magnitude scale.  So the magnitudes reported for the Harvard plates (and the Sonneberg plates) are all reliably in the modern $B$-band.  For these novae, we must convert from $B_{peak}$ to $V_{peak}$ as 
\begin{equation}
V_{peak} = B_{peak} - E(B-V) - \langle (B-V)_{peak,0} \rangle.
\end{equation}
 The last term is the intrinsic color at peak.  This color is nearly a constant from nova-to-nova, having to do with the temperature of the photosphere in the expanding nova shell being determined by the ionization temperature of hydrogen.  
 
This intrinsic color at maximum has long been realized to be approximately 0.0 mag.  The only systematic measure that I know of is from van den Bergh \& Younger (1987), who used 7 novae (after discarding two outliers) to get $+$0.23 mag, with an RMS scatter of 0.16 mag.  I can improve substantially on this old value by using 24 nova with the uncertainty in the extinction of 0.10 mag or less, as listed in GCVS order in Table 2.  The second column gives the light curve class (`LC'), with the point to be that the intrinsic color at maximum light does not depend significantly on the class of the nova.  The next two columns give my measure of the observed $B-V$ at the time of peak light in the $V$-band and my estimated $E(B-V)$ as based on methods not involving the nova colors, both in units of magnitudes.  The last column gives the derived $(B-V)_{peak,0}$.  The RMS scatter of the intrinsic colors (0.19 mag) effectively equals the average uncertainty for all 24 novae (0.18 mag), which means that the observed scatter is all measurement errors and the underlying distribution is consistent with being a single constant.  The average intrinsic color is $\langle (B-V)_{peak,0} \rangle$=$+$0.11$\pm$0.04 mag. 

\begin{table}
	\centering
	\caption{Intrinsic Color at Peak, $(B-V)_{peak,0}$}
	\begin{tabular}{llrrr} 
		\hline
		Nova & LC & $(B-V)_{peak}$    &  $E(B-V)$    &  $(B-V)_{peak,0}$  \\
		\hline
OS And	&	D	&	0.44	$\pm$	0.14	&	0.15	$\pm$	0.05	&	0.29	$\pm$	0.15	\\
V705 Cas	&	D	&	0.60	$\pm$	0.14	&	0.41	$\pm$	0.06	&	0.19	$\pm$	0.15	\\
V1065 Cen	&	P	&	0.49	$\pm$	0.28	&	0.47	$\pm$	0.05	&	0.02	$\pm$	0.28	\\
FM Cir	&	J	&	0.30	$\pm$	0.28	&	0.23	$\pm$	0.05	&	0.07	$\pm$	0.28	\\
V394 CrA	&	P	&	-0.20	$\pm$	0.14	&	0.20	$\pm$	0.10	&	-0.40	$\pm$	0.17	\\
T CrB	&	S	&	0.50	$\pm$	0.14	&	0.10	$\pm$	0.10	&	0.40	$\pm$	0.17	\\
V1500 Cyg	&	S	&	0.64	$\pm$	0.14	&	0.45	$\pm$	0.07	&	0.19	$\pm$	0.16	\\
V1668 Cyg	&	S	&	0.68	$\pm$	0.14	&	0.38	$\pm$	0.05	&	0.30	$\pm$	0.15	\\
V1974 Cyg	&	P	&	0.50	$\pm$	0.14	&	0.26	$\pm$	0.03	&	0.24	$\pm$	0.14	\\
V2491 Cyg	&	C	&	0.41	$\pm$	0.14	&	0.23	$\pm$	0.05	&	0.18	$\pm$	0.15	\\
V339 Del	&	PP	&	0.25	$\pm$	0.14	&	0.18	$\pm$	0.04	&	0.07	$\pm$	0.15	\\
V408 Lup	&	J	&	0.68	$\pm$	0.14	&	0.40	$\pm$	0.10	&	0.28	$\pm$	0.17	\\
T Pyx	&	P	&	0.30	$\pm$	0.14	&	0.25	$\pm$	0.02	&	0.05	$\pm$	0.14	\\
V4739 Sgr	&	S	&	0.55	$\pm$	0.14	&	0.46	$\pm$	0.04	&	0.09	$\pm$	0.15	\\
V5579 Sgr	&	D	&	0.73	$\pm$	0.14	&	0.72	$\pm$	0.06	&	0.01	$\pm$	0.15	\\
V5583 Sgr	&	S	&	0.42	$\pm$	0.28	&	0.34	$\pm$	0.10	&	0.08	$\pm$	0.30	\\
V5668 Sgr	&	D	&	0.40	$\pm$	0.14	&	0.21	$\pm$	0.10	&	0.19	$\pm$	0.17	\\
V6594 Sgr	&	P	&	0.36	$\pm$	0.14	&	0.30	$\pm$	0.05	&	0.06	$\pm$	0.15	\\
U Sco	&	PP	&	0.30	$\pm$	0.14	&	0.20	$\pm$	0.10	&	0.10	$\pm$	0.17	\\
NR TrA	&	J	&	-0.08	$\pm$	0.28	&	0.22	$\pm$	0.05	&	-0.30	$\pm$	0.28	\\
V382 Vel	&	S	&	0.06	$\pm$	0.14	&	0.12	$\pm$	0.03	&	-0.06	$\pm$	0.14	\\
LV Vul	&	S	&	0.95	$\pm$	0.14	&	0.57	$\pm$	0.05	&	0.38	$\pm$	0.15	\\
QV Vul	&	D	&	0.66	$\pm$	0.14	&	0.40	$\pm$	0.05	&	0.26	$\pm$	0.15	\\
V458 Vul	&	J	&	0.55	$\pm$	0.28	&	0.50	$\pm$	0.05	&	0.05	$\pm$	0.28	\\
		\hline
	\end{tabular}
\label{tab:flares}		
\end{table}

For novae that have well measured peaks in both the $B$ and $V$, equation (1) can be turned around to get a good measure of the $E(B-V)$.  This is useful for the many faint novae that are too faint to have had any useful measure of the extinction from spectroscopy.

My collected $V_{peak}$ and $E(B-V)$ values for all 402 galactic novae are presented in Table 3.  The first column is the nova in GCVS order.  The next two columns give the best $V_{peak}$ and $E(B-V)$ values.

To get the distance modulus, we also have to have an independent measure of the absolute magnitude at peak, $M_{V,peak}$.  Schaefer (2018) has already demonstrated that the various historical calibrations are poor.  So I do not know of any means to produce the best peak absolute magnitude that is any better than just taking the average.  Schaefer (2018) used {\it Gaia} DR2 parallaxes to find the average $M_{V,peak}$ is -7.0 mag, with a one-sigma scatter of 1.4 mag. 

My values of $V_{peak}$ and $E(B-V)$ are combined with the adopted $M_{V,peak}$ to produce a distance modulus, $\mu_{peak}$.  To be specific,
\begin{equation}
\mu_{peak} = V_{peak} - 3.1 E(B-V) - (-7.0).
\end{equation}
The uncertainties are calculated from the usual propagation of errors for each of the terms.  My derived $\mu_{peak}$ values are tabulated in the fourth column of Table 3 for all 402 galactic novae.

The uncertainty on the distance modulus is always $\pm$1.4 mag or somewhat larger, which is a reflection of the fact that we cannot independently know the real luminosity of the nova with any good accuracy.  This relatively poor accuracy is still useful for some purposes, including distinguishing bulge from disc novae.  For many novae, this will be the only useful specific information of their distances.  These $V_{peak}$-distances will also provide the prior information used in the Bayesian calculations from the {\it Gaia} parallaxes.  For the novae with accurate {\it Gaia} parallaxes, the $V_{peak}$-distances will have little effect on the final range of distances.  For the majority of novae with relatively poor {\it Gaia} parallaxes, the $V_{peak}$-distances will provide a prior that will force the result towards the near or far side of the range allowed from the parallaxes.

\begin{table}
	\centering
	\caption{Distance Moduli from $V_{peak}$ With $M_{V,peak}$=$-$7.0$\pm$1.4 (full table with 402 novae is in on-line supplementary material)}
	\begin{tabular}{lrrr} 
		\hline
		Nova & $V_{peak}$ (mag) & $E(B-V)$  (mag)   &  $\mu_{peak}$ (mag)  \\
		\hline

OS And	&	6.50	$\pm$	0.10	&	0.15	$\pm$	0.05	&	13.04	$\pm$	1.41	\\
CI Aql	&	9.00	$\pm$	0.10	&	0.85	$\pm$	0.30	&	13.37	$\pm$	1.68	\\
DO Aql	&	8.50	$\pm$	0.10	&	0.29	$\pm$	0.10	&	14.60	$\pm$	1.44	\\
EL Aql	&	5.32	$\pm$	0.18	&	0.97	$\pm$	0.15	&	9.31	$\pm$	1.49	\\
EY Aql	&	9.69	$\pm$	0.54	&	1.10	$\pm$	0.50	&	13.28	$\pm$	2.16	\\
...	&	&	&	\\
QV Vul	&	7.10	$\pm$	0.10	&	0.40	$\pm$	0.05	&	12.86	$\pm$	1.41	\\
V458 Vul	&	8.10	$\pm$	0.20	&	0.50	$\pm$	0.05	&	13.55	$\pm$	1.42	\\
V459 Vul	&	7.60	$\pm$	0.10	&	0.86	$\pm$	0.12	&	11.93	$\pm$	1.45	\\
V569 Vul	&	$<$16.3	$\pm$	0.20	&	3.06	$\pm$	0.50	&	$<$13.81	$\pm$	2.10	\\
V606 Vul	&	10.10	$\pm$	0.10	&	0.86	$\pm$	0.20	&	14.43	$\pm$	1.53	\\
		\hline
	\end{tabular}
\label{tab:flares}		
\end{table}

\section{DISTANCES TO BULGE NOVAE}

\begin{figure*}
	\includegraphics[width=2.0\columnwidth]{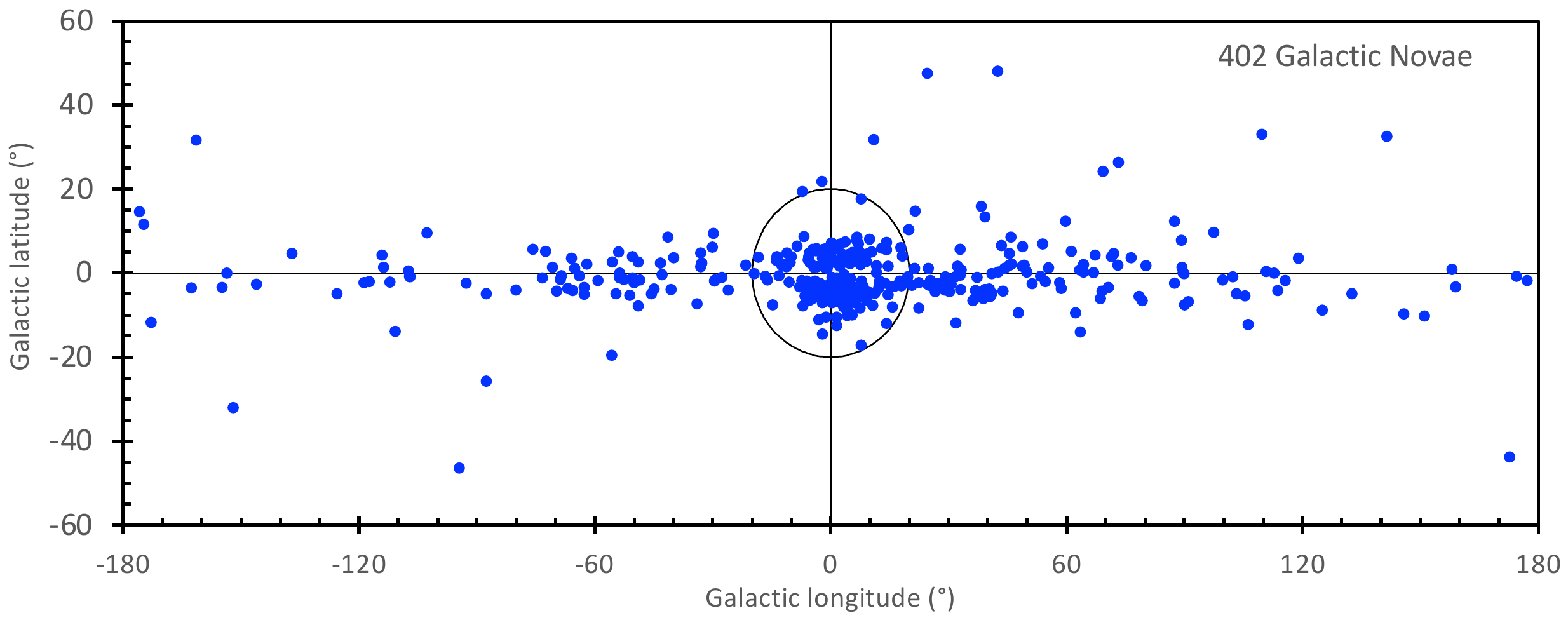}
    \caption{Galactic distribution for 402 novae.  The nova distribution is the sum of a disc population (over all galactic longitudes and mostly within 10$\degr$ of the galactic plane) and the bulge population (mostly within 9$\degr$ of the galactic centre, and all within the 20$\degr$ circle in the plot).  Nearly 41 per cent of all discovered novae are in this small clump.  With this positional information alone, we see that any nova within something like 9$\degr$ of the galactic centre has a high probability of being a bulge nova.  The gap at low galactic latitudes is prominent for the bulge novae, because they are roughly at 8000 pc distance and dimmed by very high extinction.  The low-latitude zone-of-avoidance is not significantly visible outside of the bulge.}  
\end{figure*}

\begin{figure}
	\includegraphics[width=1.0\columnwidth]{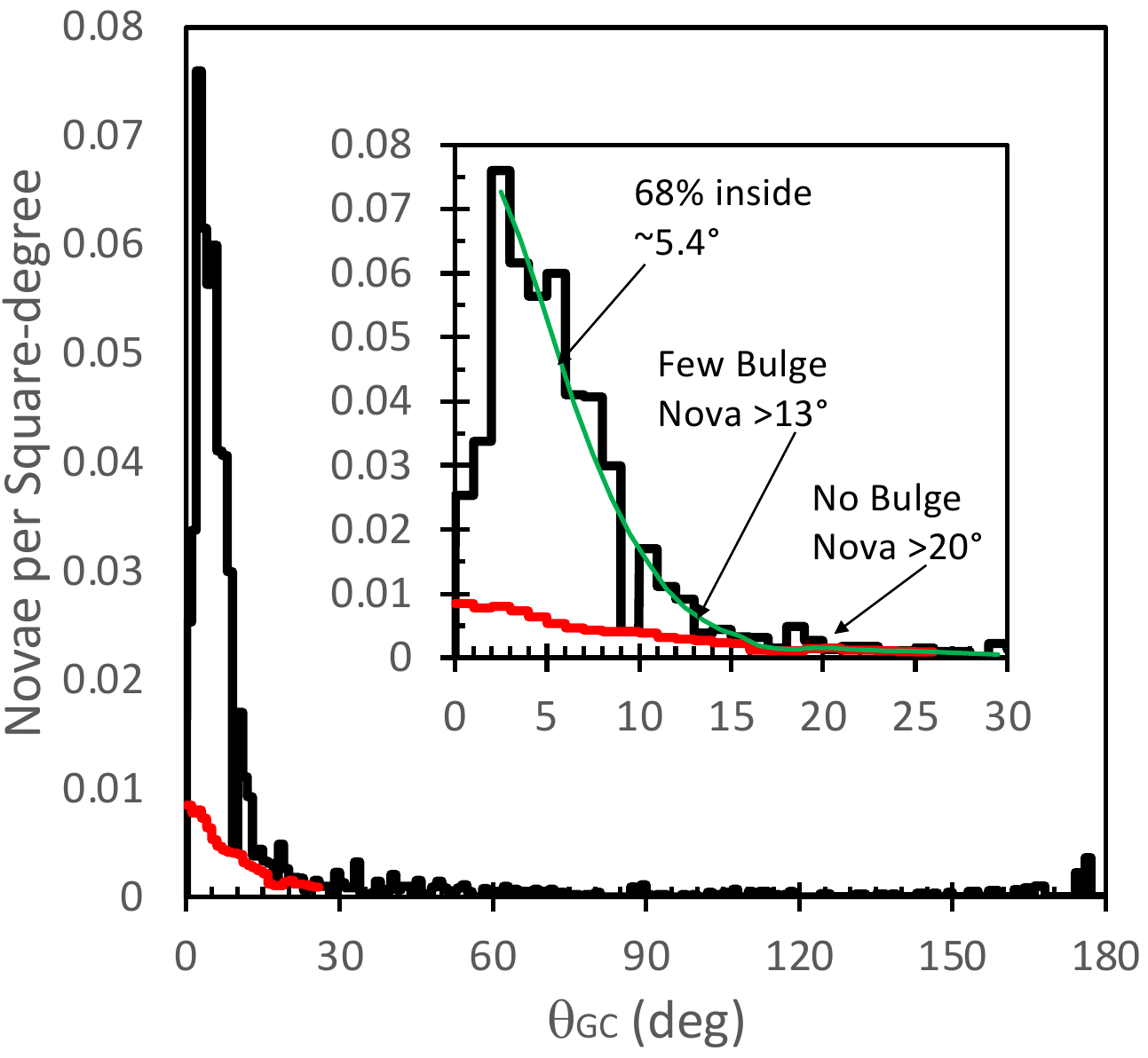}
    \caption{Distribution over angle from galactic centre for 402 novae.  One important point from this figure is to see the huge and narrow cluster of novae closely around the galactic centre, with these being the bulge population.  This cluster is shown with an expanded angular scale in the inset figure.  The best estimate for the number density of disc novae for the galactic centre region is shown by the red histogram.  In the densest part of the bulge area, the probability that a nova is in the bulge varies from roughly 80--90 per cent, as based on the $\theta_{GC}$ alone.  We see that few bulge novae appear with $\theta_{GC}$$>$13$\degr$, and no bulge novae appear more than 20$\degr$ from the galactic centre.  A description of the bulge population distribution is that of a Gaussian with a sharp drop in the innermost region due to interstellar dust.  While the small-number statistics and the dust absorption make for substantial uncertainties, the 68 per cent containment radius is roughly 5.4$\degr$.  For a distance to the galactic centre of $D_{GC}$=8000 pc, this corresponds to a characteristic radius for the bulge population of $R_{bulge}$=750 pc. }  
\end{figure}

\begin{table*}
	\centering
	\caption{Bulge/Disc Population Identification for $\theta_{GC}$$<$20$\degr$ (full table with 214 novae is in on-line supplementary material)}
	\begin{tabular}{llrrrr} 
		\hline
		Nova & $\theta_{GC}$ ($\degr$) & $\mu_{old}$  (mag)   & $\mu_{peak}$  (mag)   &  $\varpi$ (mas)   &  Population  \\
		\hline

V5854 Sgr	&	1.00	&	...				&	$<$	11.90	$\pm$	3.13	&	 	1.43	$\pm$	1.41	&	bulge	\\
V3730 Oph	&	1.57	&	...				&	$<$	12.60	$\pm$	2.10	&	 	-0.08	$\pm$	0.70	&	BULGE	\\
V4092 Sgr	&	1.77	&	...				&	$<$	14.09	$\pm$	1.69	&	...				&	bulge	\\
V2415 Sgr	&	1.77	&	...				&	$<$	16.09	$\pm$	1.72	&	...				&	bulge	\\
V5586 Sgr	&	1.79	&		12.74	$\pm$	1.01	&		11.90	$\pm$	1.69	&	...				&	DISC	\\
...	&	&	&    &    &	\\
V1313 Sco	&	18.85	&		15.25	$\pm$	0.76	&		13.71	$\pm$	1.75	&	...				&	BULGE	\\
V841 Oph	&	19.29	&	...				&		10.06	$\pm$	1.49	&	 	1.19	$\pm$	0.02	&	DISC	\\
V366 Sct	&	19.53	&	...				&	$<$	5.90	$\pm$	9.87	&	 	-0.32	$\pm$	1.53	&	DISC	\\
V1662 Sco	&	19.59	&		12.29	$\pm$	1.01	&	$<$	11.30	$\pm$	4.86	&	...				&	DISC	\\
FV Sct	&	19.68	&	...				&	$<$	14.13	$\pm$	1.92	&	 	0.48	$\pm$	0.22	&	disc	\\
		\hline
	\end{tabular}
\end{table*}

The distribution of galactic novae across the sky (see Fig. 1) shows two distinct populations.  The well-known novae (because they are bright from being close our Sun) are in the disc population, which appear close to the galactic plane all around the galactic equator.  The bulge novae (those that cluster within 20$\degr$ of the galactic centre) are clearly a distinct population.  This division of novae into disc and bulge populations is similar to the division of all Milky Way stars into disc and bulge populations.  Presumably the bulge novae are just an ordinary part of our galaxy's bulge, and represent an older population.  All of the bulge novae are relatively faint (none peaking brighter than 7.3 mag), and so there are no famous bulge novae.

With the bulge novae being faint and obscure, little work has been done on their distinct properties.  Shafter (2008) provides a review of the results.  Estimated bulge fractions ranges from some small percentage (Duerbeck 1984) up to around 75 per cent (Della Valle \& Livio 1994).  With a detailed model of structures for the bulge stars, disc stars, and interstellar extinction, Hatano et al. (1997), concluded ``the true bulge fraction of Galactic classical novae  is much closer to 1/8 than to 1/2, i.e., most Galactic classical novae are from the disc, not from the bulge".   With these large uncertainties, Shafter points out that external galaxies became the focal point for population studies.  Unfortunately, this produced contradictory studies on key properties, for example, whether of not the $M_{V,peak}$ is bimodally distributed or not, and on whether the specific rates (novae per year per unit stellar mass) are correlated with the Hubble type of the galaxy.  And we recently have the conundrum that our Milky Way has 15-out-of-20 novae with evolved companions in the bulge population (Schaefer 2022), while the Andromeda Galaxy apparently has all of such nova in the disc population (Williams et al. 2016).  It is safe to say that little is known with useable confidence about the galactic bulge nova demographics and properties.  For example, I know of no listing of galactic bulge novae, much less a compilation of their properties.

The defining property of the bulge novae is their angular distance from the galactic centre, $\theta_{GC}$.  For this, I have calculated $\theta_{GC}$ for all 402 nova, and constructed a histogram for the density of novae inside annuli of 1$\degr$ widths, in units of number-per-square-degree (Fig. 2).  This histogram shows two distinct populations, the high clump with $\theta_{GC}$$<$20$\degr$ for the bulge novae plus the low-flat component over all $\theta_{GC}$ for the disc novae.  

What is the disc component for $\theta_{GC}$$<$20$\degr$?  It is possible to try to use a detailed galactic model, but such would not have much confidence due to not knowing how nova systems relate to the other stars in the general model.  For any model or empirical results, high accuracy will not be possible, because we are dealing with small-number statistics, with there being 3 or fewer disc novae in each annulus.  There are also large systematic errors in the nova detection rates around the galactic centre, as can be seen by the large differences between negative and positive longitudes, as well as by the asymmetries between the positive and negative latitudes near the centre.  These asymmetries are likely caused by a combination of the complex history of nova search regions as well as in the galactic dust absorption.  With these asymmetries, there is no accurate means to interpolate the disc population into the galactic centre region.  A reasonable empirical method is to make identical counts in annuli for cases where all the nova coordinates are shifted by a constant in galactic longitude.  So for example, if all the longitudes are shifted by $+$40$\degr$, we have just 1 nova inside the 4.0$\degr$ to 5.0$\degr$ annulus, and if the longitudes are shifted by $-$40$\degr$ then we have 2 novae.  This is while the real sky, i.e., with zero shift, has 20 novae in the same annulus, so we can divide these up into approximately 18--19 bulge novae and 1--2 disc novae.  This preserves the statistics of the galactic latitude distribution as combined with the annuli shape, and it ensures that we are only counting disc novae.  To get useable statistics, I have evaluated the number in each of the shifted annuli for shifts of $\pm$50$\degr$, $\pm$40$\degr$, $\pm$30$\degr$, and $\pm$25$\degr$ for small annuli.  There is no significant rise in numbers from 50$\degr$ to 25$\degr$.  Averaging over these shifts and then smoothing over angles produces my best estimate for the numbers and densities of disc novae inside the galactic centre region.  This is displayed as the red line in Fig. 2.

The bulge component rises high above the disc component for $\theta_{GC}$$<$9$\degr$.  For these novae, by the galactic coordinates alone, the probability is high that any individual novae is in the bulge population.  The huge excess of bulge novae in Fig. 2 falls off sharply from 5$\degr$ to 13$\degr$.  A case can be made that the observed number of novae outside 13$\degr$ is nearly consistent with the expected numbers from the disc population.  The excess in the observed novae over the expectations for disc novae alone is just 15 novae from 13$\degr$$<$$\theta_{GC}$$<$20$\degr$, and with the uncertainties, this is barely a `3-$\sigma$' result.  The density for bulge novae outside of 20$\degr$ is certainly negligible. 


What is the characteristic radius of the bulge population, $R_{bulge}$?  I will describe this characteristic radius using terminology for a Gaussian distribution, with this usage as a reasonable description rather than a model or theory claim.  The measure of the 68 per cent containment radius has substantial uncertainties due to the observed variation around any smooth curve (see inset in Fig. 2), the uncertainties in the numbers of disc novae to subtract out as `foreground', and the sharp drop in numbers at the lowest galactic latitudes.  Nevertheless, I have fitted a Gaussian for the annuli from 2--20$\degr$.  The Gaussian sigma is 5.4$\degr$.  The formal error bar is 0.3$\degr$.  But I expect that the real error bar is larger due to systematic problems, most importantly due to the substantial changes in the nova discovery efficiency across the galactic centre region, and the real uncertainty is more like $\pm$1$\degr$.

The distance to the galactic centre, $D_{GC}$, has long been measured, and with increasing accuracy.  Reid (1993) reviewed all the prior measures and concluded that $D_{GC}$=8000$\pm$500 pc.  Malkin (2012) reviewed 52 measures of the distance from 1992--2013, yielding a final combined best distance of 7980$\pm$15$\pm$20 pc.  To all needed accuracy, I will adopt the round-number of $D_{GC}$=8000 pc.


Now, we can turn the characteristic angular size of the bulge nova population into its physical size.  This is simply $R_{bulge}$=$D_{GC}$$\sin$(5.4$\degr$), or 750 pc, with an uncertainty of roughly $\pm$130 pc.  With this, along any line of sight towards the galactic centre, every bulge nova must have a distance of 8000$\pm$750 pc, with the uncertainty containing 68 per cent of the population.  This distance corresponds to a distance modulus of 14.5$\pm$0.2 mag and corresponds to a simple parallax of 0.125$\pm$0.012 milli-arcseconds (mas).

Now, with my compilation of all 402 known novae, I can pick out a complete roster for the bulge population novae, with good confidence for each individual novae.  The first cut, based on the drop to no-excess above the disc nova population, is that bulge novae are only with $\theta_{GC}$$<$20$\degr$.  Next, I require that each candidate bulge nova have a {\it Gaia} parallax (if available) consistent with 0.125 mas.  Most bulge novae are too faint to have a {\it Gaia} parallax, so this requirement is mainly used to identify disc novae in the galactic centre region as those with a parallax too large to be consistent with the bulge distance.  Further, I require that $\mu_{peak}$ be consistent with 14.5 mag.  Most bulge novae have poor measures of $\mu_{peak}$, so this constraint is mainly useful for recognizing disc novae that are too bright to be bulge novae.  Plus, I require that any available old distance measures from the literature be consistent with $D_{GC}$, which is to say that $\mu_{old}$ is consistent with 14.5 mag.  For the 214 novae with $\theta_{GC}$$<$20$\degr$, I have placed the values of $\mu_{old}$, $\mu_{peak}$, and the {\it Gaia} parallaxes (see next Section) into Table 4.  These can be used to find consistency or inconsistency for each nova with a distance of $D_{GC}$.  

Further constraints on the nova distances can be derived from their observed extinctions and magnitudes:  {\bf (1)} If the observed extinction is greatly smaller than possible for a $\sim$8000 pc distance, then the system must be a disc nova.  For the best example, consider V4643 Sgr, with its position close to the galactic centre and $\it b$ of $-$0.34$\degr$.  The total extinction along this heavily-shrouded line-of-sight is 20.5 mag (Schlafly \& Finkbeiner 2011), so if V4643 Sgr were inside the bulge then its extinction would be $>$10 mag or so.  However, the observed extinction is 1.47$\pm$0.20 mag, and hence the nova must be much closer than the bulge and must be part of the disc population.  For a different example, V366 Sct was observed to peak at 9.94 mag, but we have no measure of its extinction.  The line-of-sight has 8.4 mag of extinction (Schlafly \& Finkbeiner 2011), so a bulge nova must have $E(B-V)$$>$4 mag or so, for $A_V$$>$12 mag.  To be in the bulge, V336 Sct would need a peak absolute magnitude more-luminous than $-$16 mag, which is certainly not a nova.  This method can demonstrate that 10 novae in the galactic centre region are certainly of the disc population.  {\bf (2)} A further constraint comes from the quiescent magnitudes with extinction corrections, $V_{q,0}$.  This uses my earlier result that novae in quiescence almost always have an absolute magnitude more-luminous than $+$5 mag (Schaefer 2018; see also Patterson et al. 2022).  So any nova appearing with $V_{q,0}$$>$19.5 is unlikely to be in the disc population.  For the specific example of V4444 Sgr, the extinction is 0.56 mag or less (Schlafly \& Finkbeiner 2011), the quiescent magnitude is $>$21 (CVCat), so the distance modulus must be $>$14.3, which demonstrates a bulge population.  Twelve novae in the galactic centre region are thus shown to be bulge novae.  {\bf (3)} A similar constraint comes because few classical novae have quiescent absolute magnitudes more-luminous than $+$3 mag.  The only exceptions are RN and those systems with subgiant and giant companions, with these being recognized by their infrared colors, even at bulge distances (Schaefer 2022).  Hence, a normal bulge nova (with a distance modulus near 14.5 mag) will appear fainter than 16.5 mag.  This can be turned around to say that a nova with $V_{q,0}$$<$16.5 is likely a disc nova.  For example, V732 Sgr is seen at $V_q$=16.9 (Mroz et al. 2015) with extinction of 0.81$\pm$0.16 mag, so $V_{q,0}$=14.4 at the distance of the bulge the absolute magnitude would have to be more-luminous than $-$0.1 mag, which is to imply that the system is in the disc population.  With this method, five novae in the galactic centre region can be assigned to the disc population.

For many novae, the evidence is clear as to whether the nova is bulge or disc.  These are indicated in the last column of Table 4 with either ``DISC" or ``BULGE".  Out of the 214 novae in Table 4, 124 have a confident population assignment.  Most disc novae will have a clear signature.  That is, they are usually bright enough to have one of the `old' methods applied, or to have their $\mu_{peak}$$\ll$14.5, or to be near enough that {\it Gaia} can detect a large parallax.  I find 14 disc novae for 0$\degr$$<$$\theta_{GC}$$<$6$\degr$, 17 disc novae from 6$\degr$ to 12$\degr$, and 4 disc novae in the range 12--18$\degr$, with this being consistent with my prior estimates.  

Nevertheless, many novae in Table 4 do not have a confident population.  A good case can be made that the remaining novae are almost all in the bulge population.  First, such novae will be those that are relatively faint and relatively far, which is to say that most of the cases will be bulge novae.  Second, roughly all of the disc novae in the bulge cluster of Fig. 2 are now identified, and the remaining novae can only be bulge sources so as to account for the huge excess at low $\theta_{GC}$.  For example, in the 0--6$\degr$ zone, 9 disc novae are expected, 14 disc novae are now confidently identified, and the zone has 83 novae, so nearly all of the 83-14=69 remaining novae must be bulge systems.  Similarly, for the 6--12$\degr$ zone, 17 disc novae are expected, 17 disc novae have been identified, so the remaining 75 novae must be almost entirely bulge novae.  In the 12--18$\degr$ zone, the 24  remaining novae (i.e., those not already identified as being disc population) must have 14 or more that are bulge systems.

I can present a good case for the probability being high for many of the novae being bulge systems, even without any good {\it Gaia} parallax or other data.  In the 0--12$\degr$ zone, the probability of such stars being a bulge system is $\gtrsim$95 per cent, simply as required to make the bulge contribution in Fig. 2 while all the disc novae are already identified.  Further, these systems have faint $V_{peak}$ and have either no {\it Gaia} counterpart or have the {\it Gaia} counterpart with a near-zero parallax, so the {\it lack} of good data are confident signs that these systems are not disc novae.  That is, any disc nova seen in the foreground of the bulge will be brighter than a bulge novae, which makes for a substantially great likelihood that {\it Gaia} will detect the counterpart and return a parallax that is significantly larger than 0.125 mas, and which makes for a substantially greater likelihood that the disc nova will have a distance measure from one of the old methods.  While any high accuracy is impossible, I estimate a probability of 10 per cent or lower that a nova with no {\it Gaia} counterpart (and no distance measures from any of the old methods) is at a distance significantly closer than 8000 pc.  The combination of these two probabilities gives $\gtrsim$99.5 per cent that each individual remaining novae is a bulge system.

This probability estimate of $\gtrsim$99.5 per cent has substantial uncertainties in its application.  First, just because the expected number of disc nova interlopers is already overfilled does not mean that some small number of additional interlopers cannot be hiding amongst the poorly observed events.  Second, a disc nova with a distance around 6000 pc could easily produce a {\it Gaia} counterpart whose parallax is too poor to be recognize as not being a bulge source, and such a nova could easily be too faint to trigger any of the various old methods to recognize its distance.  Third, in the outer zones, I have not recognized the expected number of disc novae, so likely a few more unidentified disc interlopers are hiding around the edges of the bulge.  Nevertheless, the basic probability result (that almost all of the novae in Table 4 not identified as ``DISC" must be in the bulge population) is still confident, because the bulge-identifications are required to get the high numbers in the bulge component of Fig. 2.  Therefore, the remaining non-DISC novae in Table 4 are almost all in the bulge population and are designated by `bulge', with lower case letters.

Some novae do not neatly fit into the `bulge' classification.  These cases have population assignments of `disc', `disc?', and `bulge?'.  Here are three examples to illustrate each of these population assignments:  {\bf (1)} KY Sgr has a parallax of $-$0.02$\pm$0.10 mas, clearly indicating a distance comparable to $D_{GC}$.  This nova has extinction of 0.55 mag (Shafter 1997) presumably from colors in quiescence, and 1.0$\pm$0.5 mag from my fits to the spectral energy distribution (Schaefer 2022).  The extinction along the entire line-of-sight is 3.29 mag (Schlafly \& Finkbeiner 2011), while most of this must come from distances closer than the bulge (because with galactic latitude -1.72$\degr$ the line-of-sight past the bulge is already far outside the disc).  The discrepancy arises in trying to understand how KY Sgr can suffer much less than the maximum extinction, unless its distance is significantly closer than the 8000 pc.  I do not know how to resolve this discrepancy, so I am labeling this system as `bulge?'.  {\bf (2)} Some of the novae have apparently contradictory information.  For example, V630 Sgr has a parallax of 0.14$\pm$0.11 mas pointing to a bulge distance (with parallax near 0.125 mas), yet has a well-observed peak with $V_{peak,0}$=2.2 (implying a peak absolute magnitude of $-$12.3 for a galactic centre distance).  In this case, I reconcile the two measures to a distance of 4000 pc, corresponding to a one-sigma deviation from the best-estimate parallax and corresponding to a peak absolute magnitude of $-$10.8 mag at the extreme end of the known distribution.   I am labeling this as a `disc?' nova.  {\bf (3)} The `disc' novae are those with the measures pointing to the disc population, yet for which the evidence does not produce high confidence.  FV Sct has a parallax of 0.48$\pm$0.22 mas, for which the simple calculation gives a distance of 2000 pc, indicating a disc population.  However, a parallax of 0.125 mas for bulge sources is only 1.6-sigma off the best-estimate parallax, so the parallax alone does not yield a decisive answer.  The only other useful measure, $\mu_{peak}$$<$14.13$\pm$1.92, is ambiguous, where the large error bars allow for both usual disc and bulge distances.  I am labeling this case as `disc' for the most likely population, but the lower case lettering indicates that this result is not of high confidence.

A minor question concerns the two novae that are inside globular clusters.  Globular clusters are certainly in the older bulge (or halo) population, but the dynamical histories of these two novae are likely greatly different from the other bulge novae.  I will dodge this question by assigning these to a separate ``GlobC" population type.  

In the end, many population assignments have good confidence as based on measured distances (the 38 `DISC' and 86 `BULGE' novae), most of the remainder have assignments with good confidence due to the requirement to get the large numbers of bulge novae as an excess above the disc component (the 75 `bulge' novae), while a small fraction have assignments that are likely correct but individually of lesser confidence (the 4 `bulge?', 9 `disc?', and 2 `disc' novae).  Of the 214 novae in Table 4, 165 are identified as being in the bulge.  The bulge fraction for all 402 known galactic novae is 41 per cent.  This is the first list of bulge novae that I am aware of.  And now, this list can be used for appropriate priors in the distance calculation.

\section{GAIA DR3 PARALLAXES}

\begin{table}
	\centering
	\caption{{\it Gaia} Parallaxes and Magnitudes for 215 Galactic Novae (full table with 215 novae is in on-line supplementary material)}
	\begin{tabular}{lllll} 
		\hline
		Nova & $\varpi$ (mas)  & $b$  (mag)   & $g$  (mag)   &  $r$ (mag) \\
		\hline

OS And	&	0.21	$\pm$	0.11	&	18.29	&	18.14	&	17.82	\\
CI Aql	&	0.38	$\pm$	0.04	&	16.43	&	15.79	&	14.94	\\
DO Aql	&	0.35	$\pm$	0.18	&	18.10	&	18.15	&	17.04	\\
EL Aql	&	0.16	$\pm$	0.08	&	18.10	&	16.59	&	15.42	\\
EY Aql	&	-0.67	$\pm$	0.70	&	21.05	&	20.20	&	19.12	\\
...	&	&	&    &   	\\
NQ Vul	&	0.84	$\pm$	0.07	&	18.02	&	17.29	&	16.43	\\
PW Vul	&	0.47	$\pm$	0.08	&	17.85	&	17.66	&	17.21	\\
QU Vul	&	1.11	$\pm$	0.31	&	19.70	&	19.56	&	18.80	\\
QV Vul	&	0.38	$\pm$	0.18	&	18.87	&	18.56	&	17.86	\\
V458 Vul	&	0.19	$\pm$	0.11	&	18.30	&	18.10	&	17.80	\\
		\hline
	\end{tabular}
\end{table}

On 2022 July 13, the {\it Gaia} Team released the much-anticipated Data Release 3 (DR3), with its impressive collection of parallaxes for stars down to roughly 19th magnitude and fainter from all over the sky.  For the nova community, we finally have a collection of accurate and reliable distances for a large number of novae.

{\it Gaia} DR3 is part of a progression with great improvements in accuracy and number.  Take the case of the brightest old novae, V603 Aql.  Reported parallaxes are $\varpi$=4.011$\pm$0.137 from $HST$ (Harrison et al. 2013), 2.92$\pm$0.54 mas from {\it Gaia} DR1 (Ramsay et al. 2017), 3.191$\pm$0.069 mas from {\it Gaia} DR2 (Schaefer 2018), 3.11$\pm$0.03 mas from {\it Gaia} EDR3, and now 3.106$\pm$0.034 mas from {\it Gaia} DR3.  The number of novae with accurate parallaxes progressed from 4 for $HST$, to 3 for DR1, 41 for DR2, and now 74 for DR3.

The {\it Gaia} spacecraft is described in Gaia Collaboration et al. (2016), while DR3 is described in Gaia Collaboration et al. (2022).  DR3 data can be accessed from a European Space Agency website\footnote{https://gea.esac.esa.int/archive/}.

A bureaucratic mode for extracting the {\it Gaia} parallaxes would be to run the standard search program and record the parallax for the star nearest to the nova position.  But this mindless approach would be bad for the majority of novae.  One reason is that many dozens of novae have no identified quiescent counterpart, and no accurate position, so that the parallax for the {\it Gaia} star nearest to the cataloged nova position has likely no connection to the nova parallax.  Another reason is that $\sim$10 per cent of the faint novae have well-known counterparts that are more than one arc-second away from the catalogued positions, so a blind search would likely turn up the wrong answer.  Further, many novae have a crowded field, such that many stars are close to the known nova position, so it is unclear as to which {\it Gaia} star is the real counterpart.  A final problem is that many novae have the counterpart being faint, below the {\it Gaia} threshold, while the blind positional-coincidence method will then often take the meaningless parallax for a foreground star.

A proper use of the {\it Gaia} database for nova requires that some sort of a positive connection be made between the nova counterpart and the {\it Gaia} star.  This connection can be made when the counterpart is known to be isolated to substantially fainter than the known counterpart magnitude, and the {\it Gaia} star is similarly isolated and of roughly the same magnitude as the counterpart.  Further evidence might be that the {\it Gaia} star is relatively blue in color, or perhaps noted as a variable star.  For the not-simple cases, I have used the {\it Gaia} catalog to construct a small star chart showing the relative positions and magnitudes, for direct comparison with a deep image on which the counterpart is positively identified.  Excellent finder charts appear in the SMARTS catalog (Walter et al. 2012), the Duerbeck catalog (Duerbeck 1987), and the CVCat (Downes \& Shara 2001).  For purposes of this paper, I have included the {\it Gaia} parallax only when I can convince myself that the evidence is good for the {\it Gaia} star being the nova counterpart.

I have exhaustively searched for a confident {\it Gaia} counterpart for each of the 402 known galactic novae.  In the majority of the cases, this has entailed making a star chart of the {\it Gaia} stars for comparison with picture of the real sky with tick marks identifying the counterpart or its location.  For many other novae, the search consisted of seeing that {\it Gaia} recorded no stars within 2 arc-seconds of the nova position, and confirming in some other source that the {\it Gaia} position is the correct one.  In 20 cases, {\it Gaia} DR3 reports on the nova counterpart, but no parallax is included.  I have found a total of 195 nova parallaxes with DR3.  From this, only 74 galactic novae have `good' accuracy.  Here `good' accuracy is taken to be where the uncertainty on the parallax is $<$0.30 times the parallax ($\sigma_{\varpi}$ $<$0.30$\varpi$), with this being the limit beyond which the prior contributes more to the final distance than does the new parallax information.  For parallaxes with uncertainties from 0.30 to 1.0 times the parallax, I will label these measures as ``poor'', for which my list has 66 novae.  When $\sigma_{\varpi}$$>$$\varpi$, there is still marginal information, effectively just defining a lower limit on the nova distance, for which I have 55 novae from the DR3 catalog.  Out of these near-zero parallaxes, 36 are actually negative parallaxes.

All of the 215 {\it Gaia} counterparts have their parallaxes and magnitudes reported in Table 5.  The first column gives the nova name in the correct GCVS order, the second column gives the {\it Gaia} DR3 parallax in mas, and the last three columns give the {\it Gaia} magnitudes in the $b$, $g$, and $r$ bands.  To get distances in parsecs, these parallaxes must be combined with the priors from $\mu_{old}$ (see Table 1), $\mu_{peak}$ (see Table 3), and the population assignment (see Table 4).

\section{USING ALL THE INFORMATION TO GET THE BEST DISTANCES}

The simple and traditional calculation of distance ($D$ in parsecs) from a parallax ($\varpi$ in mas with uncertainty $\sigma_{\varpi}$) is that $D=1000/\varpi$ with an uncertainty of $\sigma_D=D*(\sigma_{\varpi}/\varpi$).  This is fine when the fractional error of the parallax is small.  When the fraction error ($\it f$=$\sigma_{\varpi}/\varpi$) is not small, substantial problems arise (Bailer-Jones 2015).  For $\it f$$\gtrsim$0.30, the calculated probability distribution can become bimodal (where the results are telling us more about the prior information than about any new input), so the 30 per cent threshold is fine for recognizing parallaxes with good accuracy.  To account for the non-linearities, the solution is to use a Bayesian calculation, as recommended by the {\it Gaia} Team (Luri et al. 2018).  Detailed and helpful explanations and examples are given in Bailer-Jones (2015).

The $\it unnormalized$ probability for the nova being at distance $D$, given the measured $\varpi$ and $\sigma_{\varpi}$, is
\begin{equation}
P(D|\varpi, \sigma_{\varpi}) = P(\varpi|D, \sigma_{\varpi})~P(D).
\end{equation}
The first factor on the right side of the equality is the probability that a nova at distance D will return a measured parallax of $\varpi$.  The second factor on the right side is the `prior', which encapsulates all previous knowledge of the nova distance.  The first factor is
\begin{equation}
P(\varpi|D, \sigma_{\varpi}) = \frac{1}{\sigma_{\varpi}} exp \left[\frac{-1}{2\sigma_{\varpi}^2}(\varpi - \frac{1}{D})^2 \right].
\end{equation}
Throughout, the probability distributions can be multiplied by a normalization factor (i.e., the function integrated over all $D$) so that the integral of all the final probabilities is unity.

The prior information included in $P(D)$ can come from a variety of sources.  Here, I will be including the information of the galactic position, the distance modulus from the old methods, and the distance modulus from $V_{peak}$.  These are multiplicative, so
\begin{equation}
P(D) = P_{gal}(D|\ell,{\it b}) \times P_{old}(D|\mu_{old}) \times P_{peak}(D|\mu_{peak}).
\end{equation}
The inclusion of these factors depends on the prior information for the nova, before and independent of the {\it Gaia} parallax.

The $P_{gal}$ factor is representing the number of novae along a sightline as a function of distance, and this is the product of the nova space density as a function of distance times the volume of space in the expanding cone of the sightline.  The $P_{gal}$ factor depends on whether the nova is in the bulge or disc population.  For a disc population,  
\begin{equation}
P_{gal}(D) = D^2 \times \exp \left[ \frac{-D \sin({|\it b}|)}{\langle |Z| \rangle} \right]  \times \exp \left[ \frac{-R_{GC}}{h_R} \right].
\end{equation}
The $D^2$ factor represents the greater volume of space as we go to larger and larger distances.  The first exponential factor represents the fact that disc populations have their number density decreasing exponentially with their vertical distance from the centre of the galactic plane, $Z$=$D$$\sin({\it b})$.  Disc novae are observed to have an average value of $|Z|$, i.e., the exponential scale height $\langle |Z| \rangle$, for which I measure 140 pc (see Section 7.1).  Along the line of sight towards a galactic latitude of $\it b$, the space number density of disc novae will fall off exponentially with distance.  The exponential term involving the galactocentric distance, $R_{GC}$, is to represent the known exponential fall-off in star density of our Milky Way, with the scale length, $h_R$, roughly equal to 3000 pc.  The value of $R_{GC}$ is just a function of $D$, $D_{GC}$, and the galactic longitude $\ell$.  This representation of our galaxy does not incorporate any spiral arms or a bar in the centre.  The exponential in $R_{GC}$ has little effect on any distances, other than to cap the distances for poorly-observed disc novae at low galactic latitude.  This formulation has suppressed constant factors, as these all are absorbed into a normalization factor.  This is the `exponentially decreasing space density' (EDSD) prior (see equation 18 of Bailer-Jones 2015) recommended by the {\it Gaia} Team (Luri et al. 2018).  

For a bulge novae, we can represent their radial distribution as a Gaussian along any line of sight with a central distance of $D_{GC}$ and a one-sigma scatter of $R_{bulge}$.  As a useable approximation for bulge novae, our prior information includes a Gaussian in $D$;
\begin{equation}
P_{gal}(D) = D^2 \times \exp \left[ - \frac{1}{2} \frac{(D-D_{GC})^2}{R_{bulge}^2} \right].
\end{equation}
Again, the $D^2$ factor covers the expanding volume at greater distances.  I adopt $D_{GC}$=8000 pc and $R_{bulge}$=750 pc.  All of the galactic novae will have the $P_{gal}$ prior from either equation 6 or 7.

The prior information from the old methods and from $V_{peak}$ have the distance moduli distributed as a Gaussian.  For the prior based on the old methods,
\begin{equation}
P_{old}(D|\mu_{old}) = \exp \left[ - \frac{1}{2} \frac{(5 Log[D] - 5 - \mu_{old})^2}{\sigma_{old}^2} \right].
\end{equation}
Many of the old measures from extinction are {\it lower limits}, and this can be handled by making the $\sigma_{old}$ to be very large for any distance larger than the limit.  In the closer direction, the quoted $\sigma_{old}$ should be used as a reflection of the uncertainty in the value of the limit.  Similarly, the prior for the distance information from $V_{peak}$ will be
\begin{equation}
P_{peak}(D|\mu_{peak}) = \exp \left[ - \frac{1}{2} \frac{(5 Log[D] - 5 - \mu_{peak})^2}{\sigma_{peak}^2} \right].
\end{equation}
Many novae only have faint limits on $V_{peak}$, or upper limits on $\mu_{peak}$.  This should be handled by setting $\sigma_{peak}$ to the tabulated value for distances larger than the tabulated limit (to represent the uncertainty in the limit) or to a very large number for distances less than the tabulated limit (to represent the limit on the distance).  With this, we can calculate the overall probability distribution for each nova, as a function of D and observed parameters.

This Bayesian formulation has a strong advantage that it provides a natural means to collect together all available information on $D$, where the old methods and the $V_{peak}$ are correctly combined with the new {\it Gaia} parallaxes.  The results are the best possible distances, taking into account all inputs.  This is what I am presenting as the ``Overall Best Distances" in this paper's title.

The calculated probability as a function of distance (from equation 3) should be normalized to an area under the curve of unity.  (For display purposes, the probability curves are best normalized so that the peak probability is at unity, as then the shapes are not misperceived due to radical changes in the heights.)  The resultant distribution functions always have an asymmetric and non-standard shape.  (That is, the distributions are {\it not} Gaussians in distance.)  These distributions record all the best information on $D$.  However, presenting these distributions is awkward, and comparisons are hard.  The solution is to extract from each distribution the most probable $D$ and some range for representing the uncertainties.  Following the recommendation of Bailer-Jones, I am using the mode of the probability distribution as the best estimator of $D$.  That is, the $D$ values I list will be for those at the maximum probability of the probability distribution.  Further, to represent the reasonable range of $D$, I will not use the $\pm$ notation (e.g., 1210$^{+390}_{-220}$ pc), but will instead quote a range of $D$ (e.g., 990--1600 pc).  The range I chose to describe is that which contains the central 68 per cent of the probability distribution.  The reason for choosing 68 per cent is because astronomers are practiced at knowing the meaning, whereas other choices would only be confusing.

\begin{figure}
	\includegraphics[width=1.01\columnwidth]{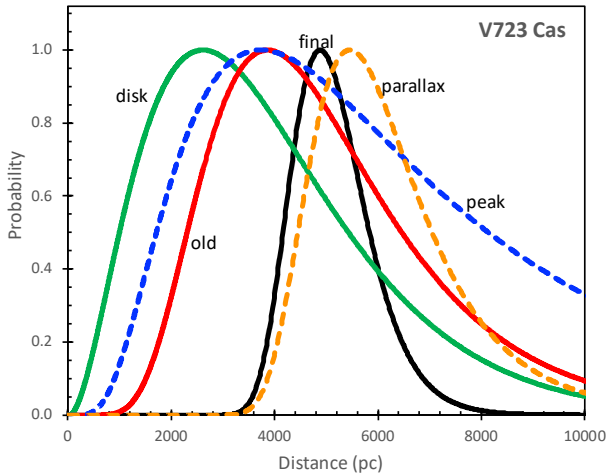}
    \caption{The probability distributions for V723 Cas.  The final probability distribution is the unnormalized product of four separate probability distributions, each representing information on the distance to the nova.  The $P_{gal}$ distribution (solid green curve) represents the knowledge about the real distribution of novae in our galaxy, in this case for the space density of disc novae falling off exponentially with distance from the centre plane of our galaxy.  The $P_{old}$ distribution (solid red curve) represents the old measure of the expansion parallax as 3860 pc, with a real uncertainty of near 1690 pc.  The $P_{peak}$ distribution (dashed blue curve) is based on the observed peak magnitude of 7.10$\pm$0.10, the observed $E(B-V)$=0.40$\pm$0.10, and the adopted average for all novae of $M_{V,peak}$ of $-$7.0 mag.  The broad width of this distribution is because the one-sigma on the peak absolute magnitude being $\pm$1.4 mag.  The $P(\varpi|D, \sigma_{\varpi})$ distribution (orange dashed curve) is from the {\it Gaia} DR3 parallax of 0.183$\pm$0.035 mas.  The $P(D|\varpi, \sigma_{\varpi})$ distribution (solid black curve) is the final best estimate of the probability of V723 Cas being at the given distance.  This final curve is simply the product of the four input curves as displayed.  (For display purposes only, the distributions have all been normalized so that the maximum probability is unity.)  The narrowest of the four distributions is from the {\it Gaia} parallax, so the other three distributions serve only to push the distance somewhat closer.  In the end, the best estimate of the distance to V723 Cas is that the highest point in the final curve is at 4880 pc, while the central 68 per cent of probability is inside the range 4390--5910 pc.}
\end{figure}

To illustrate this calculation (see Fig. 3), let me take the case of V723 Cas, as a typical nova with a moderate fraction error in the parallax ($\it f$=0.19), plus an old measure of the expansion parallax and a well measured peak magnitude.  In Cassiopeia, the nova is certainly in the disc population and with a galactic latitude of $\it b$ at -8.81$\degr$.  Just given this galactic position, distances closer than 600 pc or so are unlikely simply because there is little volume inside this distance, while distances farther than 6000 pc or so are unlikely due to the nova then being required to be very far below the galactic centre plane.  So just from the galactic position alone, the expected distance can be narrowed down to a range that is a factor of 10$\times$ in size.  For the old measures, we have an expansion parallax with a distance of 3860 pc, but the real uncertainty is much larger than published, as represented by the broad red curve in Fig. 3.  For the $P_{peak}$ information, the best estimate distance is 3730 pc as based on the peak magnitude alone.  However, the $\pm$1.4 mag uncertainty on our adopted absolute means that the $P_{peak}$ distributions will always be very broad.    The {\it Gaia} parallax provides the best information, but even this is only accurate at the 19 per cent level.   When all four of these distributions are multiplied together, the three non-parallax factors only serve to push the final product to the near-side of the distribution allowed by the parallax.  The parallax alone is pointing to distances centred on 5450 pc, but all three of the other sources of information are pointing to nearer distances, and these provide a weak pull on the parallax to produce a final distance of 4880 pc as the most probable value.

\begin{table*}
	\centering
	\caption{Distances and Properties of All 402 Known Galactic Novae (Supplementary Material has additional columns for $\ell$, {\it b}, $\Theta_{GC}$, $E(B-V)$, $t_2$, and $\varpi$)}
\begin{tabular}{lllrlllllll} 
		\hline
		Nova & Year  & LC($t_3$)  & $V_{peak}$   &  Spec.  &   FWHM   &   $P$   &   Pop.   &   $D$ (68\% range)  &   $M_{V}$    &  Properties$^a$ \\
		\hline

OS And	&	1986.9	&	D(23)	&	6.50	$\pm$	0.10	&	...	&	1740	&	...	&	DISC	&	4152	~(	3549	 --	5349	)	&	-7.06	&	...	\\
CI Aql	&	RN	&	P(32)	&	9.00	$\pm$	0.10	&	...	&	2250	&	0.61836	&	DISC	&	2761	~(	2555	 --	3143	)	&	-5.84	&	SubG, Ecl, $\Delta P$, $\dot{P}$	\\
DO Aql	&	1925.7	&	F(900)	&	8.50	$\pm$	0.10	&	Fe {\rm II}	&	...	&	0.16776	&	DISC	&	3151	~(	2571	 --	4826	)	&	-4.89	&	Ecl	\\
EL Aql	&	1927.6	&	J(25)	&	5.32	$\pm$	0.18	&	...	&	1230	&	...	&	DISC	&	4419	~(	3781	 --	6450	)	&	-10.91	&	...	\\
EY Aql	&	1926.7	&	S(40)	&	9.69	$\pm$	0.54	&	...	&	...	&	...	&	DISC	&	5819	~(	4214	 --	9613	)	&	...	&	...	\\
V356 Aql	&	1936.7	&	J(140)	&	7.00	$\pm$	0.10	&	Fe {\rm II}	&	1140	&	0.42651	&	DISC	&	3243	~(	2668	 --	5725	)	&	...	&	...	\\
V368 Aql	&	1936.8	&	S(42)	&	5.30	$\pm$	0.20	&	Fe {\rm II}	&	...	&	0.69051	&	DISC	&	2532	~(	2324	 --	2948	)	&	-8.58	&	SubG, Ecl	\\
\hline
	\end{tabular}
\end{table*}

\begin{table*}
	\centering
	\contcaption{Distances and Properties of All 402 Known Galactic Novae}
	\label{tab:continued}
	\begin{tabular}{lllrlllllll} 
		\hline
		Nova & Year  & LC($t_3$)  & $V_{peak}$   &  Spec.  &   FWHM   &   $P$   &   Pop.   &   $D$ (68\% range)  &   $M_{V}$    &  Properties$^a$ \\
		\hline

IV Cep	&	1971.5	&	S(37)	&	7.00	$\pm$	0.40	&	Fe {\rm II}	&	...	&	...	&	DISC	&	6463	~(	5632	 --	9009	)	&	-9.07	&	...	\\
V809 Cep	&	2013.2	&	D(36)	&	11.18	$\pm$	0.10	&	Fe {\rm II}	&	500	&	...	&	DISC	&	6162	~(	5051	 --	10594	)	&	...	&	...	\\
V962 Cep	&	2014.3	&	S(42)	&	11.00	$\pm$	0.10	&	Fe {\rm II}	&	2000	&	...	&	DISC	&	5545	~(	4606	 --	7262	)	&	-5.63	&	...	\\
HV Cet	&	~2008.8	&	...	&	$<$14.2	$\pm$	0.20	&	Ne	&	2900	&	1.7718	&	DISC	&	380	~(	260	 --	881	)	&	...	&	SubG, PreERise	\\
RR Cha	&	1953.3	&	...	&	$<$6.7	$\pm$	0.25	&	...	&	...	&	0.1401	&	DISC	&	1860	~(	1615	 --	2628	)	&	...	&	Ecl	\\
X Cir	&	1927.0	&	...	&	$<$6.1	$\pm$	0.25	&	...	&	1420	&	0.15446	&	DISC	&	3267	~(	2144	 --	5718	)	&	...	&	Ecl	\\
AR Cir	&	1906.1	&	J(330)	&	8.29	$\pm$	0.45	&	...	&	...	&	0.2143	&	DISC	&	1456	~(	1306	 --	1842	)	&	...	&	...	\\
\hline
	\end{tabular}
\end{table*}

\begin{table*}
	\centering
	\contcaption{Distances and Properties of All 402 Known Galactic Novae}
	\label{tab:continued}
	\begin{tabular}{lllrlllllll} 
		\hline
		Nova & Year  & LC($t_3$)  & $V_{peak}$   &  Spec.  &   FWHM   &   $P$   &   Pop.   &   $D$ (68\% range)  &   $M_{V}$    &  Properties$^a$ \\
		\hline

IM Nor	&	RN	&	P(80)	&	8.50	$\pm$	0.10	&	...	&	2380	&	0.20717	&	DISC	&	4312	~(	3461	 --	6326	)	&	-7.15	&	Ecl, $\dot{P}$	\\
V341 Nor	&	1983.7	&	S(14)	&	9.40	$\pm$	0.50	&	...	&	3000	&	...	&	DISC	&	7436	~(	5476	 --	11209	)	&	...	&	...	\\
V382 Nor	&	2005.2	&	S(18)	&	8.70	$\pm$	0.40	&	Fe {\rm II}	&	1850	&	...	&	DISC	&	5615	~(	4341	 --	8119	)	&	-7.68	&	...	\\
V390 Nor	&	2007.5	&	J	&	9.10	$\pm$	0.30	&	Fe {\rm II}	&	...	&	0.17133	&	DISC	&	4174	~(	3337	 --	6010	)	&	-6.79	&	...	\\
V555 Nor	&	2016.3	&	S(110)	&	12.70	$\pm$	0.40	&	He/N	&	2140	&	...	&	DISC	&	7561	~(	5476	 --	11720	)	&	...	&	...	\\
V556 Nor	&	2018.9	&	S(9)	&	10.10	$\pm$	0.40	&	Fe {\rm II}	&	2800	&	...	&	DISC	&	5840	~(	4028	 --	9161	)	&	...	&	...	\\
RS Oph	&	RN	&	P(14)	&	4.80	$\pm$	0.10	&	...	&	3930	&	453.6	&	DISC	&	2710	~(	2575	 --	2908	)	&	-9.63	&	RG, $\gamma$, PostEdip	\\
\hline
	\end{tabular}
\end{table*}

\begin{table*}
	\centering
	\contcaption{Distances and Properties of All 402 Known Galactic Novae}
	\label{tab:continued}
	\begin{tabular}{lllrlllllll} 
		\hline
		Nova & Year  & LC($t_3$)  & $V_{peak}$   &  Spec.  &   FWHM   &   $P$   &   Pop.   &   $D$ (68\% range)  &   $M_{V}$    &  Properties$^a$ \\
		\hline

V351 Pup	&	1992.0	&	P(26)	&	6.40	$\pm$	0.10	&	Fe {\rm II}, Ne	&	...	&	0.1182	&	DISC	&	3869	~(	3077	 --	5996	)	&	-8.55	&	Shell	\\
V445 Pup	&	2000.9	&	D(240)	&	8.60	$\pm$	0.10	&	Helium	&	...	&	...	&	DISC	&	6272	~(	5026	 --	9026	)	&	-6.97	&	Shell	\\
V574 Pup	&	2004.9	&	S(27)	&	7.00	$\pm$	0.10	&	Fe {\rm II}, Ne	&	2830	&	...	&	DISC	&	4567	~(	3744	 --	7174	)	&	-7.85	&	...	\\
V597 Pup	&	2007.6	&	S(6)	&	6.20	$\pm$	0.50	&	Hybrid	&	1800	&	0.11119	&	DISC	&	5851	~(	4682	 --	9384	)	&	-8.57	&	Ecl, IP?	\\
V598 Pup	&	2007.6	&	...	&	$<$10.2	$\pm$	0.20	&	He/N	&	2100	&	0.16287	&	DISC	&	1844	~(	1755	 --	1965	)	&	...	&	NonOrbP	\\
T Pyx	&	RN	&	P(62)	&	6.40	$\pm$	0.10	&	Hybrid 	&	5350	&	0.07623	&	DISC	&	3599	~(	3369	 --	3931	)	&	-7.16	&	InGap, Ecl, $\Delta P$, $\dot{P}$, 	\\
 & 	& 	&	 	&	 	&	 &	 	& 	&	 	&	 	&	 PreERise, V1500, Sh	\\
\hline
	\end{tabular}
\end{table*}

\begin{table*}
	\centering
	\contcaption{Distances and Properties of All 402 Known Galactic Novae}
	\label{tab:continued}
	\begin{tabular}{lllrlllllll} 
		\hline
		Nova & Year  & LC($t_3$)  & $V_{peak}$   &  Spec.  &   FWHM   &   $P$   &   Pop.   &   $D$ (68\% range)  &   $M_{V}$    &  Properties$^a$ \\
		\hline

V4160 Sgr	&	1991.6	&	S(3)	&	7.00	$\pm$	0.10	&	He/N, Ne	&	4500	&	...	&	DISC	&	2602	~(	2062	 --	4091	)	&	-6.07	&	...	\\
V4169 Sgr	&	1992.5	&	S(36)	&	7.90	$\pm$	0.10	&	Fe {\rm II}	&	...	&	...	&	DISC	&	2939	~(	2322	 --	4561	)	&	-5.49	&	...	\\
V4171 Sgr	&	1992.8	&	S(20)	&	6.60	$\pm$	0.20	&	Fe {\rm II}	&	1200	&	...	&	bulge	&	7946	~(	7206	 --	8697	)	&	-9.76	&	...	\\
V4327 Sgr	&	1993.6	&	S(30)	&	8.10	$\pm$	0.30	&	Hybrid	&	2400	&	...	&	bulge	&	8092	~(	7359	 --	8837	)	&	-7.62	&	...	\\
V4338 Sgr	&	1990.1	&	...	&	8.00	$\pm$	0.50	&	...	&	400	&	29.48192	&	BULGE	&	8117	~(	7399	 --	8853	)	&	-8.41	&	RG	\\
V4361 Sgr	&	1996.5	&	...	&	$<$10.3	$\pm$	0.20	&	Fe {\rm II}	&	...	&	...	&	DISC	&	6845	~(	4332	 --	9474	)	&	...	&	...	\\
V4362 Sgr	&	1994.4	&	D(121)	&	7.80	$\pm$	0.20	&	Fe {\rm II}	&	450	&	...	&	bulge	&	8074	~(	7344	 --	8828	)	&	...	&	...	\\
\hline
	\end{tabular}
\end{table*}

\begin{table*}
	\centering
	\contcaption{Distances and Properties of All 402 Known Galactic Novae}
	\label{tab:continued}
	\begin{tabular}{lllrlllllll} 
		\hline
		Nova & Year  & LC($t_3$)  & $V_{peak}$   &  Spec.  &   FWHM   &   $P$   &   Pop.   &   $D$ (68\% range)  &   $M_{V}$    &  Properties$^a$ \\
		\hline

V697 Sco	&	1941.2	&	...	&	$<$9.8	$\pm$	0.25	&	...	&	1550	&	1.26716	&	DISC	&	870	~(	780	 --	1257	)	&	...	&	SubG, Ecl	\\
V707 Sco	&	1922.5	&	...	&	9.14	$\pm$	0.45	&	...	&	...	&	...	&	bulge	&	8119	~(	7377	 --	8855	)	&	-7.42	&	...	\\
V711 Sco	&	1906.3	&	S	&	9.08	$\pm$	0.18	&	...	&	...	&	...	&	bulge	&	8128	~(	7400	 --	8875	)	&	-7.05	&	...	\\
V719 Sco	&	1950.5	&	...	&	8.55	$\pm$	0.32	&	Fe {\rm II}	&	1800	&	0.43639	&	bulge	&	8057	~(	7315	 --	8801	)	&	...	&	...	\\
V720 Sco	&	1950.6	&	...	&	$<$6.9	$\pm$	0.25	&	Fe {\rm II}	&	1800	&	...	&	BULGE	&	7997	~(	7256	 --	8743	)	&	...	&	...	\\
V721 Sco	&	1950.7	&	...	&	$<$9.2	$\pm$	0.29	&	...	&	...	&	...	&	DISC	&	1857	~(	1748	 --	2028	)	&	...	&	...	\\
V722 Sco	&	1952.2	&	...	&	$<$8.7	$\pm$	0.29	&	...	&	...	&	...	&	bulge	&	8063	~(	7331	 --	8812	)	&	...	&	...	\\
\hline
	\end{tabular}
\end{table*}

\begin{table*}
	\centering
	\contcaption{Distances and Properties of All 402 Known Galactic Novae}
	\label{tab:continued}
	\begin{tabular}{lllrlllllll} 
		\hline
		Nova & Year  & LC($t_3$)  & $V_{peak}$   &  Spec.  &   FWHM   &   $P$   &   Pop.   &   $D$ (68\% range)  &   $M_{V}$    &  Properties$^a$ \\
		\hline

CT Ser	&	1948.3	&	...	&	$<$7.7	$\pm$	0.23	&	He/N	&	...	&	0.195	&	DISC	&	2551	~(	2330	 --	2904	)	&	...	&	...	\\
DZ Ser	&	1960.6	&	...	&	$<$13.1	$\pm$	0.36	&	He/N	&	...	&	...	&	BULGE	&	8142	~(	7396	 --	8883	)	&	...	&	...	\\
FH Ser	&	1970.1	&	D(62)	&	4.50	$\pm$	0.10	&	Fe {\rm II}	&	...	&	...	&	DISC	&	1024	~(	970	 --	1103	)	&	-7.41	&	Shell	\\
LW Ser	&	1978.2	&	D(52)	&	8.30	$\pm$	0.10	&	Fe {\rm II}	&	...	&	...	&	bulge	&	7661	~(	6933	 --	8412	)	&	-7.33	&	...	\\
MU Ser	&	1983.1	&	...	&	7.70	$\pm$	0.40	&	He/N	&	1230	&	...	&	BULGE	&	7262	~(	6503	 --	8025	)	&	-7.85	&	...	\\
V378 Ser	&	2005.2	&	J(93)	&	11.60	$\pm$	0.40	&	Fe {\rm II}	&	1100	&	...	&	BULGE	&	8238	~(	7500	 --	8969	)	&	-5.27	&	...	\\
V556 Ser	&	2014.0	&	...	&	$<$12.3	$\pm$	0.20	&	Fe {\rm II}	&	1145	&	...	&	BULGE	&	8166	~(	7420	 --	8903	)	&	...	&	...	\\
\hline
	\end{tabular}
	
\begin{flushleft}	
\
$^a$ Novae below and inside the nova Period Gap from 0.071--0.111 days are denoted with {\bf $<$Gap} and {\bf inGap}.  Novae with subgiant companion stars (roughly 0.6$<$$P$$<$10 days) and red giant companion stars (roughly $P$$>$10 days) are marked as {\bf SubG} and {\bf RG}.  Novae whose light curves display eclipses are denoted with {\bf Ecl}, display dwarf novae eruptions with {\bf DN}, display inexplicable pre-eruption rises with {\bf PreERise}, display post-eruption dips with {\bf PostEDip}, display coherent and stable periods that are certainly not orbital or rotational  with {\bf NonOrbP}, display very large amplitude irregular variability in quiescence with {\bf LargeAmpVar}, and display various types of inexplicable rebrightenings are notated with {\bf Rebrighten}.  {\bf V1500} indicates the mysterious V1500 Cyg stars where the post-eruption brightness many decades after the end of the eruption remains at least 10$\times$ brighter than the pre-eruption level.  {\bf InPNeb}; V458 Vul is the only nova to be at the centre of an observed ordinary planetary nebula.  Novae visible with $\gamma$-ray emission by the  {\it Fermi} spacecraft are marked {\bf $\gamma$}.  Novae for which I have measured the steady period change between eruptions are marked with {\bf $\dot{P}$}, while those for which I have measured the change in the orbital period sharply across a nova eruption are marked with {\bf $\Delta P$}.  V1500 Cyg is the only nova known to be an asynchronous polar ({\bf AP}).  Novae inside host globular clusters are listed with {\bf in M14} or {\bf in M80}.  Binaries that are either confidently or likely Intermediate Polars ({\bf IP} or {\bf IP?}) are identified as in the catalogue of K. Mukai (https://asd.gsfc.nasa.gov/Koji.Mukai/iphome/catalog/alpha.html).  Startlingly, V2487 Oph suffers the most-energetic and the most-frequent {\bf Superflares} out of the many classes of stars that display bright optical flares caused by stellar magnetic reconnection.
\\

\end{flushleft}	
	
\end{table*}

The resultant probability distributions are often roughly symmetric and similar to Gaussians in shape.  But many novae have distributions that are substantially asymmetric.  The most extreme cases are for FM Sgr and V1175 Sgr, their sharp peaks in the distribution (i.e., the mode) ares outside the middle 68 per cent of the distribution. 

I have calculated the best-estimate $D$ values and the middle 68 per cent probability ranges for all 402 novae.  These are presented in Table 6.  I have a total of 220 novae with good accuracy in their distances.  (For this, I take a `good' distance to be one where half the quoted distance range is $\leq$0.3$D$.)  Out of these 219 novae with good distances, 52 are disc novae, with the primary basis for the accuracy being their {\it Gaia} parallaxes.  The set of novae with good distances includes all 165 bulge novae.  (These include only 20 novae with $\sigma_{\varpi}$$<$0.10 mas, so the distance accuracy is almost entirely due to the identification as a bulge novae, with this being confident and reliable in all but a few marginal cases.)  In addition the two novae in globular clusters have accurate {\it Gaia} parallaxes to the cluster.   In the end, as the primary goal of my program, I have 220 novae (out of the 402 known galactic novae) that have accurate distances.

Table 6 also lists many of the fundamental measured properties for all 402 galactic novae.  The first column gives the nova names, in the required GCVS order.  The second column gives the year of the peak, or gives ``RN'' for recurrent novae.  The third column gives the light curve classification according to the divisions of Strope et al. (2010), consisting of one or two letters (S, P, PP, O, C, D, J, and F) plus the $t_3$ time (the number of days from the nova peak until the last time it fades below 3.0 mag under the peak) in parentheses.  All of these classifications are from my own work with the original light curves for this paper.  The fourth column gives the peak V-band magnitude, $V_{peak}$ and its one-sigma uncertainty.  These magnitudes are taken from my analysis of the original light curves, with the B-magnitudes converted to V-magnitudes, and care taken to identify when the peak was reliably detected.  The fifth column gives the spectral class, with the main classes being Fe {\rm II}, He/N, and hybrid, while neon-novae are indicated with an ``Ne''.  The next column gives the FWHM of the H$\alpha$ line in units of km/s, fairly early in the eruption.  These data are often heterogeneous, with the only line width information coming from other hydrogen lines, or only the FWZI is quoted, or late in the eruption.  The next column lists the 156 known orbital periods, $P$ in days, as collected in Schaefer (2022), with this including my 49 new orbital periods.  The eighth column gives my assignments for the nova population, as taken from Section 4.  The ninth column gives my derived distances, $D$ in parsecs, along with the range of the central 68 per cent of the probability distribution, with this column being the primary product of this paper.  The tenth column gives the absolute magnitude at peak in the V-band, $M_{V,peak}$.  These values are all calculated from the tabulated $V_{peak}$, $E(B-V)$, and $D$, with full extinction correction.  The print copy of this table only shows the $M_{V,peak}$ values for those with uncertainties better than 1.0 mag.  The on-line Supplementary Material has this column containing the measured values for all the novae, plus their formal one-sigma error bars.  The last column gives a listing of the various unusual properties for each nova, with the notations keyed in a footnote to the table.

I have added a number of additional columns of measured properties, with these appearing only in the on-line Supplementary Material.  Three of the columns give the galactic coordinates $\ell$, {\it b}, and $\Theta_{GC}$, all in units of degrees.  The next column gives my final values for $E(B-V)$ and its one sigma error bars, in units of magnitudes, for all 402 novae.  Many of these are taken from \"{O}zd\"{o}nmez et al. (2018), or from subsequent literature, or from my own light curve analyses, or from prudent estimates based on the upper limits from Schlafly \& Finkbeiner (2011).  The next column gives the $t_2$ value, which is the number of days over which the nova light curve declines from its highest peak until the last time the nova fades by more than 2.0 mags from the peak.  The last extra column gives the {\it Gaia} DR3 parallax and its one sigma uncertainty, as in Table 5.

Table 6 is intended to be an exhaustive summary of all the known properties of all 402 known galactic novae.  The one line of information gives a comprehensive summary of all its known properties, and each line reveals the `personality' of each nova.  This full table can be used for looking up properties of individual novae, for standardizing the prior results from the long history of measuring novae, and for demographic studies of a wide variety of nova properties.

\section{APPLICATIONS}

\subsection{Nova Disc Scale Height}

Duerbeck (1984) collects nine measures of the exponential scale height for novae, with values ranging from 72--440 pc, and then went on to use 20 novae to derive his own scale height of 125 pc.  The height of each nova above the central plane of our galaxy, $Z$, is $D$$\sin$($\it b$).  My histograms of the distances from the galactic plane, $|Z|$, shows that all sets of disc novae have a sharp drop in numbers from $|Z|$=0, and the distributions are close to exponential.  The distribution can be accurately represented by the exponential scale height, which is the same as the average, $\langle |Z| \rangle$.


The derived $\langle |Z| \rangle$ will vary with the selection of disc novae included in the average.  If the set for averaging includes novae with large uncertainty in $D$, then $\langle |Z| \rangle$ will systematically grow in size to be comparable to the error bars.  So the set of novae for averaging should be restricted to only the best-measured novae with uncertainties substantially smaller than 150 pc or so.  But if we make the set too small, then the scale height will be uncertain due to small-number statistics.  Further, some sort of volume-limited sample is needed to minimize selection effects.  To avoid edge effects in this volume, we need to use an infinite cylinder, centred on the Sun, perpendicular to the galactic plane.  Further, I have not used any novae with $\theta_{GC}$$<$20$\degr$, so as to avoid any possible confusion with bulge novae.

I have adopted a middle set, with a cylinder radius of 2000 pc and requiring that the uncertainty in $Z$ be less than 30 pc.  This results in a set of 34 disc novae with good distances.  This set has $\langle |Z| \rangle$=140 pc.  The uncertainty is defined by the reasonable variations for inclusion in the averaging set, with this being $\pm$10 pc.  So I conclude that the novae are distributed exponentially above and below the galactic plane, with an exponential scale height of 140$\pm$10 pc.  This $\langle |Z| \rangle$=140 pc value is used in equation (6) to calculate the final nova distances, as in Section 6 and Table 6.

\subsection{Absolute Magnitude At Peak Light In The V-band}

\begin{figure}
	\includegraphics[width=1.0\columnwidth]{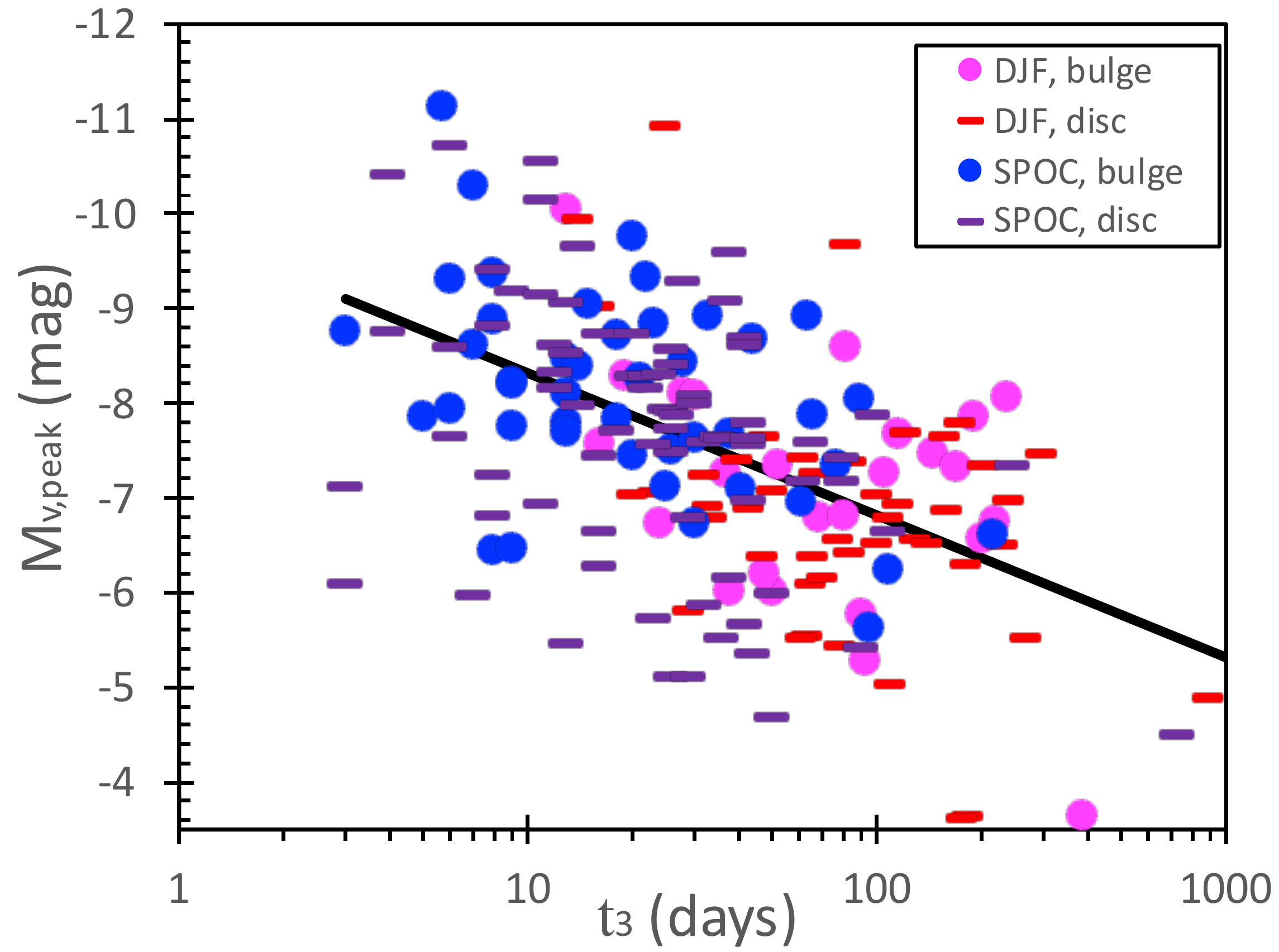}
    \caption{Absolute magnitude at peak ($M_{V,peak}$) versus decline rate ($t_3$).  This plot shows the MMRD for 192 novae with measured $t_3$ and the uncertainty in $M_{V,peak}$ of less than 1.0 mag.  We see a significant correlation, $M_{V,peak} = -7.6 + 1.5 \log(t_3 / 30)$, as marked by the slanted line.  But we also see a huge scatter, such that the `correlation' just looks like a round blob.  The only reason that the correlation is significant is because I have 192 novae.  All the prior samples testing the MMRD have had much fewer novae, and thus the many samples have not shown any significant MMRD relation, while one prior sample has randomly happened to show a poor MMRD.  The MMRD shown here supersedes all prior galactic results.  The conclusion is that the MMRD does really exist, but the huge scatter dominates over the effect and thus the relation is not useable for any purpose that I can think of.}  
\end{figure}

What is the average $M_{V,peak}$?  Schaefer (2018) used {\it Gaia} DR2 data for novae with accurate parallaxes to yield an average of $-$7.0 with an RMS scatter of 1.4 mag.  Now, with the 213 novae in Table 6, the average is $-$7.45 mag with an RMS scatter of 1.33 mag.  Various subsets give similar values, with a variation over 0.2 mag or so, for example, the 15 novae with error bars $<$0.30 mag have an average of $-$7.51 mag with an RMS scatter of 1.35 mag.  For the full set of 213 novae, the top end of the luminosity distribution has 8 novae evenly spread over the range $-$10.0$>$$M_{V,peak}$$>$$-$11.13.  These include the famous V1500 Cyg at $-$10.38, and the extreme case is the bulge nova V3661 Oph at $-$11.13.  At the low-luminosity end, 6 novae span $-$3.62$>$$M_{V,peak}$$>$$-$5.0.  The extreme case is V972 Oph at $-$3.62 mag.  Three of the six cases have flat-topped F-class light curves (DO Aql, V1310 Sgr, and BT Mon), pointing to this class being systematically low-energy eruptions.  For all of these extreme cases, I see no indications that the $M_{V,peak}$ values have any large measurement or systematic errors past what is expressed in the stated error bars.  So it appears that the range of the nova phenomenon extends from at least $-$5.0 to $-$10.0.

These D and $M_{V,peak}$ values were derived with an assumed prior for peaks of -7.0$\pm$1.4 mag.  Should this be updated to the new average of $-$7.45$\pm$1.33 mag?  Or should we change the $P_{peak}$ priors to represent the various correlations with light curve and spectral classes (see below)?  The answer is clearly `no', for two strong reasons.  First, the Bayesian method requires the use of {\it prior} information, so it is incorrect to feed the calculated results back into the calculation.  Second, doing so would make for little difference anyway.  That is, the adopted error bar of $\pm$1.4 mag is already so weak that any change to the adopted $M_{V,peak}$ will lead to changes in $D$ that are greatly smaller than the quoted error bars.

Now with 213 good measures, I can search for correlations between $M_{V,peak}$ and any of many other properties listed in Table 6.  Most properties have no significant correlation with the peak luminosity, including the various spectral classes (e.g., Fe {\rm II}, He/N, and Neon), the shell expansion velocity (as measured by the FWHM of the Balmer lines), the status as a RN, and the galactic population (disc or bulge).  I find only two properties with significant correlations.  

The first correlation is with the light curve classification.  The S-class, P-class, O-class, and C-class novae are for the high-energy novae (i.e., with large FWHM, He/N spectral class, high white dwarf masses, and fast declines), and these are grouped together here.  The D-, J-, and F-classes are for low-energy novae, and are grouped together here.  With 124 SPOC novae, the average is $-$7.75$\pm$0.12.  With 75 DJF novae, the average is $-$6.95$\pm$0.15.  So just by looking at the shape of the light curve, we can identify whether the nova peaks at relatively high-or-low luminosity.

The second correlation is that $M_{V,peak}$ is a function of the decline rate of the light curve (expressed as $t_3$ or $t_2$).  Specifically, the 92 fast novae (say, with $t_3$$<$30 days) have $\langle M_{V,peak} \rangle$ equal to $-$8.13$\pm$0.13, while the 20 slow novae (with $t_3$$>$100 days) have $-$6.80$\pm$0.25.  That is a 1.33 mag difference, which is a factor of 3.4$\times$ in peak luminosity.  Fig. 4 shows a plot of $M_{V,peak}$ versus $t_3$, and we see a significant trend, despite the huge scatter making the data points looking like a round blob.  The same trend with the same large scatter is seen when only using novae with the uncertainty in $M_{V,peak}$ that is $<$0.5 mag.  This trend is identical for SPOC versus JDF light curve classes, and identical for bulge versus disc novae.  A least square fit gives the relation $M_{V,peak} = -7.6 + 1.5 \log(t_3 / 30)$, for $t_3$ in units of days.  The existence of this correlation is significant at the 8.6-sigma level.  But the scatter about this relation is huge, so the correlation is significant only because I am using 192 novae.  The scatter has an RMS of 1.0 mag on top of the measurement errors.  The relevant figure is that the disagreement from model to observation has the 68 per cent range of 2.4 mag and the total range of 6.2 mag.  That is, the 1-sigma uncertainty in using this relation results in a range of luminosity by a factor of 10$\times$.  With the very large scatter around the relation, there is little utility for physical models of any specific system.    With this large scatter, the use of this relation is a negligible improvement over using a simple average of $-$7.0$\pm$1.4 mag.

This relation fits into a long history of the so-called `Maximum Magnitude versus Rate of Decline' (MMRD).  The MMRD was first put forth for novae in the Andromeda Galaxy (Hubble 1929), with follow-ups, extensions, and refinements by many workers from the 1950s until recently, with the first theoretical explanation by Shara (1981).  MMRD relations have been presented with many formulations, involving magnitudes at various times in the eruption, involving B- and V-magnitudes, involving decline rates over 2 and 3 magnitudes, and involving logarithms, broken logs, and arc-tangents.  The best expression of the MMRD was in Downes \& Duerbeck (2000), with no improvements since then.  Historically, the MMRD has been heavily used, largely because it was the only way to estimate distances to most novae.  The MMRD has always been know to be poor, due to the huge scatter in the relation.  Starting with Kasliwal et al. (2011), the relation fell into disrepute when the novae in the galaxies M31, M81, M82, NGC 2403, and NGC 891 did {\it not} display anything like the MMRD.  They also showed that the RNe do not obey the MMRD, nor do the latest theory models of Yaron et al. (2005).  Then Shara et al. (2017b) showed that the novae in M87 do {\it not} follow the MMRD.  Then Schaefer (2018) used {\it Gaia} DR2 parallaxes to show that 26 galactic novae in the `Gold' sample followed a shifted MMRD with huge scatter, while the 13 and 23 galactic novae in the `Silver' and `Bronze' samples do not show anything like any MMRD.  The Gold sample is largely the same as the Downes \& Duerbeck sample, so we only really have just one sample of novae that poorly follows the MMRD.  In stark opposition, we have the MMRD being denied for two other galactic samples, the RNe, theory models, and by novae in six external galaxies.

Now, with 192 novae and the best distances, my confirmation of the MMRD supersedes all the prior results for our Milky Way.  Presumably the old samples either showed or denied the poor MMRD due to the randomness of small samples out of a highly scattered distribution.  That is, the Downes \& Duerbeck sample displayed a poor MMRD by the random happenstance of the novae going into their sample, while other samples had the happenstance to have the scatter being too large to allow the MMRD to be visible.  The point is that the MMRD is now seen to be real, even though the scatter is so large as to debilitate the use of this relation.  We are now left with the question as to why the Milky Way has a poor MMRD, while the six external galaxies do not show it at all.  I think that the reconciliation is that the searches in each external galaxy produces a relatively small number of novae, and the small numbers with huge scatter will always make for no significant detection of the MMRD.  That is, the huge scatter means that the MMRD can only be confidently detected when a large number of novae are included (like 192 galactic novae in this paper), whereas no MMRD can be significant when relatively small numbers of novae are included.

In the end, we are left with $\langle M_{V,peak} \rangle$=$-$7.45$\pm$1.33 mag being nearly as good a description of the data as $M_{V,peak} = -7.6 + 1.5 \log(t_3 / 30)$.  The MMRD is definitely a real effect for galactic novae, and likely for external galaxies , despite the large scatter.  The theory explanation is as given by Shara (1981), with both $M_{V,peak}$ and $t_3$ directly tied to the white dwarf mass.  Despite now having a significant existence, the MMRD has so much scatter that it is useless for evaluating the distance to any particular novae.

\subsection{Bulge Versus Disc Novae}

\begin{table}
	\centering
	\caption{Comparing Bulge and Disc Populations}
	\begin{tabular}{lll}
		\hline
		Property & Bulge Novae & Disc Novae \\
		\hline
$\langle M_V \rangle$ (mag)	&	$-$7.73 (26, $\pm$1.28)	&	$-$7.51 (45, $\pm$1.32)	\\
Median FWHM (km/s)	&	1500 (93, $\pm$1700)	&	1670 (116, $\pm$1220)	\\
Median $t_3$ (days)	&	28 (79, $\pm$67)	&	38 (163, $\pm$110)	\\
SPOC fraction	&	0.62 (51 / 82)	&	0.61 (105 / 173)	\\
Fe {\rm II} fraction	&	0.80 (77 / 96)	&	0.72 (106 / 147)	\\
He/N fraction	&	0.14 (13 / 96)	&	0.18 (27 / 147)	\\
Hybrid fraction	&	0.06 (6 / 96)	&	0.09 (13 /147)	\\
Neon fraction	&	0.08 (8 / 100)	&	0.19 (29 / 151)	\\
Helium fraction	&	0.00 (0 / 100)	&	0.01 (1 / 151)	\\
Subgiant fraction	&	0.22 (8 / 37)	&	0.18 (21 / 115)	\\
Red giant fraction	&	0.35 (13 / 37)	&	0.05 (6 / 115)	\\
		\hline
	\end{tabular}
\end{table}

Now, I have a confident list of all known bulge novae and disc novae, plus an exhaustive list of their primary known properties.  So I can systematically make a comprehensive study of the two populations.  

For this, I have used the 161 novae identified as `BULGE' or `bulge' and the 224 novae identified as `DISC'.   The results are in Table 7.  Each line presents a property for which either a fraction or an average can be derived.  The parenthetical numbers are either the numbers of novae in the fraction, or the number of novae in the average along with the overall RMS of the population.  The population differences in $\langle M_V \rangle$, the median FWHM of the velocity width of the Balmer lines, and the median $t_3$ decline speed are all within one-sigma.  Similarly, the fraction of novae that are SPOC (i.e., with light curve classes S, P, O, or C), as well as the fractions for the various spectral classes, are indistinguishable between bulge and disc novae.  Further, the fraction of novae with subgiant companion stars (i.e., those with 0.6$<$$P$$<$10 days) is the same between the populations.

I can find only one property for which the bulge and disc novae differ, and that is the red giant fraction.  The nova systems with red giant companions are easily recognized, even out past $D_{GC}$, by their infrared flux whose spectral energy distribution is a warm blackbody, shining above the usual blue accretion disc flux (Schaefer 2022).  Their orbital periods are from 10 days out to near 2440 days.  The bulge novae have a red giant fraction of 35$\pm$8 per cent, while the disc novae have a red giant fraction of 5$\pm$2 per cent.  This is a highly significant difference.  A similar result was found by Schaefer (2022).

So the only difference between the bulge and disc populations is that the bulge has a high red giant fraction.  The first-glance explanation is that the bulge novae are older, on average, than disc novae, so the nascent wide binaries formed with a white dwarf will have longer time for the companion to evolve to its red giant stage and come into contact.  But my simplistic modeling suggests that this explanation cannot be fine tuned to get such a large difference.  Further, such an explanation has a hard time allowing for the subgiant fractions to be equal between the two populations.  Another mystery arises because the novae with red giant companions in the Andromeda Galaxy are consistent with being entirely in the {\it disc} population (Williams et al. 2016).  I am not aware of any selection effect or systematic problem that can account for the large differences in the red giant fractions.  So I am left with no useable explanation for the Milky Way novae having the systems with red giants concentrated in the bulge population.

The bulge novae are neither faster or slower, neither brighter or dimmer than disc novae, nor are the bulge novae different in light curve or spectral class.  Other than the red giant fraction, the bulge novae are indistinguishable from the disc novae.

\subsection{RS Oph Is Powered By Roche Lobe Overflow}

The issue of the distance to RS Oph determines the nature of the accretion in the binary (Schaefer 2009; Wynn 2008).  For distances $<$2000 pc, the companion must be substantially smaller than its Roche lobe, so the accretion can only come by the white dwarf capturing some fraction of the red giant's stellar wind.  For distances over $\gtrsim$2700 pc, the companion must be so large as to fill its Roche lobe, and the accretion would be entirely by Roche lobe overflow.

Historically, the first widely seen distance estimates were those of Hjellming et al. (1986) and Cassatella et al. (1985), with extinction distances of 1600 pc.  Most later papers explicitly cited these papers, and 1600 pc became the default by repetition.  Then modelers took results assumed from this distance (the system having wind accretion) to calculate models, and then derived distances comparable to 1600 pc, a circular conclusion.  In an uncritical review of RS Oph distances up until 2006, Barry et al. (2008) merely collected the many published distance measures and took the median, to get 1400 pc.  The legacy of this band-wagon-effect (Schaefer 2008) has continued to the current time, with the most recent RS Oph publication (Cheung et al. 2022) adopting 1600 pc as explicitly attributed to the Hjellming paper.

The problems for this vote-on-the-distance result are that the distances reported by both Hjellming{\footnote{The mistake in Hjellming et al. (1986) was to adopt a linear calibration for their measure of extinction, with this only being reasonable for a line of site that keeps exactly in the galactic plane.  RS Oph has a galactic latitude of $\it b$=$+$10.37$\degr$, so the line of sight to RS Oph will be 300 pc out of the plane for a distance of 1600 pc and far outside almost all of the galaxy's extinction.  That is, Hjellming's calibration relation is greatly wrong, and the distance estimate is greatly wrong.  In reality, the observed extinction is close to the maximum extinction along the entire line of sight through our galaxy, and this demonstrates only that the RS Oph distance has a lower limit of something like 1000 pc, beyond which the line of sight has no significant galactic absorption.} and Cassatella{\footnote{Cassatella et al. (1986) noted that they could not find any absorption at a radial velocity corresponding to the Carina Arm of our Milky Way (at a distance of 2000 pc), and concluded that RS Oph must be closer than 2000 pc.  The error of Cassatella is in not realizing that at 2000 pc and with $\it b$=10.37$\degr$, the line of sight passes 370 pc above the galactic centre plane.  With the line-of-sight passing far above all of the gas and dust in the Carina Arm, there is no limit on the distance from this basis.} are certainly and greatly mistaken.  Further errors in the old distance estimates to RS Oph are listed in Schaefer (2009), while the real uncertainties in many of the old papers are greatly larger than reported (Schaefer 2009; 2018) so that the results are too poor to be useful.  Further, with the observed red giant wind velocities, the fraction of the stellar wind captured by the white dwarf must be greatly too small to  allow for RN eruptions every 9--27 years (Schaefer 2009).

Into this poor history, we now have a good parallax from {\it Gaia} DR3 of 0.373	$\pm$0.023 mas.  With a fractional error this small, the simple calculation returns a good distance, which is 2680$\pm$165 pc.  With the Bayesian calculation and the priors, RS Oph has a best-estimate distance of 2710 pc, with the central 68 per cent range 2575--2908 pc.  Now, {\it Gaia} has returned a decisive answer, with RS Oph near 2710 pc, and the companion star is the same size as its Roche lobe, so the accretion in the system is entirely by Roche lobe overflow.

\section{ACKNOWLEDGEMENTS}

In this massive data mining program, the real heroes are the many thousands of observers contributing their observations, plus the hundreds of workers supporting and archiving the analysis and data.  Of particular importance are the builders, operators, analysts, and archivists of the {\it Gaia} mission.  (This work has made use of data from the European Space Agency (ESA) mission
{\it Gaia} (\url{https://www.cosmos.esa.int/gaia}), processed by the {\it Gaia}
Data Processing and Analysis Consortium (DPAC,
\url{https://www.cosmos.esa.int/web/gaia/dpac/consortium}).)

\section{DATA AVAILABILITY}

The {\it Gaia} data are publicly available on-line.  The nova light curve and eruption data are from the cited literature (and references therein), and the public databases of the AAVSO.  The resultant measures and analysis are completely presented in the Tables.


{}

\bsp	
\label{lastpage}
\end{document}